\def\apj{Astrophys. J.}
\documentclass[a4paper,11pt]{article}
\usepackage{jcappub}
\usepackage{epsfig}
\usepackage{graphicx}
\usepackage{color}

\begin{document}
\title{Critical points of the cosmic velocity field and the uncertainties  in the value of the Hubble constant}
\author{Hao Liu$^{*\diamond}$, Roya Mohayaee${^{\dagger *}}$ and Pavel Naselsky$^*$ . }

\affiliation{
$^*$ Niels Bohr Institute and DISCOVERY center, Blegdamsvej 17, 2100 Copenhagen, {\O}, Denmark\\
$^\dagger$ CNRS,UPMC, Institut d'astrophysique de Paris (IAP),  98bis Boulevard Arago, 75014 Paris, France\\
$^\diamond$ Key laboratory of particle and astrophysics, Institute of High Energy Physics, CAS, China\\
}

\newcommand{\bi}[1]{\mbox{\boldmath $#1$}}
\newcommand{\nc}{\newcommand}
\newcommand{\beq}{\begin{equation}}
\newcommand{\eeq}{\end{equation}}
\newcommand{\be}{\begin{eqnarray}}
\newcommand{\ee}{\end{eqnarray}}
\newcommand{\num}{\nu_\mu}
\newcommand{\nue}{\nu_e}
\newcommand{\nut}{\nu_\tau}
\newcommand{\nus}{\nu_s}
\newcommand{\mnus}{M_s}
\newcommand{\taus}{\tau_{\nu_s}}
\newcommand{\nnt}{n_{\nu_\tau}}
\newcommand{\rnt}{\rho_{\nu_\tau}}
\newcommand{\mnt}{m_{\nu_\tau}}
\newcommand{\tnt}{\tau_{\nu_\tau}}
\newcommand{\rar}{\rightarrow}
\newcommand{\lar}{\leftarrow}
\newcommand{\lrar}{\leftrightarrow}
\newcommand{\dm}{\delta m^2}
\newcommand{\mpl}{m_{Pl}}
\newcommand{\mbh}{M_{BH}}
\newcommand{\nbh}{n_{BH}}
\newcommand{\crit}{{\rm crit}}
\newcommand{\ini}{{\rm in}}
\newcommand{\cmb}{{\rm cmb}}
\newcommand{\rec}{{\rm rec}}
\newcommand{\mnras}{MNRAS}
\newcommand{\apjs}{ApJS}
\newcommand{\aap}{A\&A}
\newcommand{\aj}{AJ}
\newcommand{\Odm}{\Omega_{\rm dm}}
\newcommand{\Ob}{\Omega_{\rm b}}
\newcommand{\Om}{\Omega_{\rm m}}
\newcommand{\nb}{n_{\rm b}}
\newcommand{\red}{\textcolor[rgb]{1.00,0.00,0.00}}
\def\simlt{\lesssim}
\def\simgt{\gtrsim}
\def\Cl{C_{\ell}}
\def\out{{\rm out}}
\def\in{{\rm in}}
\def\mean{{\rm mean}}
\def\zrec{z_{\rm rec}}
\def\zreio{z_{\rm reion}}
\def\wmap{{\it WMAP}}
\def\planck{{\it Planck}}

%\begin{abstract}
\abstract{
The existence of critical points for the peculiar velocity field is a natural feature of the correlated vector field. These points appear at the junctions of velocity domains with
different orientations of their averaged velocity vectors. Since peculiar velocities are the important cause of the scatter in
the Hubble expansion rate,  we propose that a more precise determination of
the Hubble constant can be made by restricting analysis to a subsample of observational data containing only the zones around the critical points of the peculiar velocity field, associated with voids and saddle points.
On large-scales the critical points, where the first derivative of the gravitational potential vanishes, can easily be identified using the density field and classified by the behavior of the Hessian of the gravitational potential.  We use high-resolution N-body simulations to show that these regions are stable in time and hence are excellent tracers of the initial conditions. Furthermore, we show that the variance of
the Hubble flow can be substantially minimized by restricting observations to the subsample of such regions of vanishing velocity instead of aiming at increasing the statistics by averaging indiscriminately using the full data sets, as is the common approach.
%This could help to reconcile the CMB-measured value of the Hubble constant with that obtained from the local measurements.
}
%\end{abstract}

\maketitle

\section{Introduction}
%------------------------------

Recent Cosmic Microwave Background (CMB) observations, especially the Planck mission, have yield precise values for the cosmological parameters \citep{planck2014,planck2015,2013ApJS..208...19H}.
Discrepancies between the values of certain cosmological parameters obtained from the CMB versus other observations {\it e.g.} galaxy surveys, have however frequently surfaced.
A notable discrepancy is that between the planck-determined value of the Hubble parameter and that found from  local measurements.
The low value of the
Hubble constant, $H_0=(68\pm 0.7)\,$km/s/Mpc, determined from the WMAP and BAO seems robust and has been confirmed by the Planck 2015 mission which gives $H_0=(67.27\pm 0.66)\,$km/s/Mpc
(see \citep{planck2014,planck2015}).
On the other hand, direct measurement of the Hubble parameter, using distance ladders such as the Cepheids   has persistently favored a higher value of $ 73.8 \pm 2.4~{\rm km/s/Mpc}$ for $H_0$ \citep{riess2011,riess2016} which is about $2.5\sigma$ at variance with the CMB-based value. Various remedies have been suggested such as metallicity-dependent luminosity which could lower the Cepheid-determined value of the Hubble parameter  ~\citep{efstathiou2014} or addition of sterile and massive neutrinos \cite{wyman2014}. Furthermore, parameter determination solely from CMB without the use of other data such as the baryonic acoustic oscillations have also been shown to be flawed especially at high $l$ values (see {\it e.g.} \citep{addison2016}).

A plausible reason for this discrepancy could be the sample variance of the peculiar velocity field \citep{wang98,hannestad,2014PhRvL.112v1301B}.
%The effect of \red{such} variance which increases with decreasing ``size'' of the surveys, is also expected to largely account for the disagreement between different values of the Hubble parameter.
The reason is that the contribution of peculiar velocities (see Sec~\ref{sec:Peculiar velocities and uncertainties of the Hubble flow} for the definition of peculiar velocity) to the cosmic flow, which can be ignored at high Redshifts is dominant locally and especially for surveys of limited depth.
Interestingly, direct measurement of the Hubble parameter based on masers which targets only sources of very small peculiar velocities favors a low value of the Hubble parameter: $H_0 = 68.9 \pm 7.1$~\cite{reid2013}, in good agreement with the CMB-based value.

Peculiar velocities are excellent tracers of the inhomogeneities in the distribution of total mass and hence measure of deviation from the pure Hubble flow which can only exist in a homogeneous and isotropic Universe.
Subsequently, they are nuisances for the determination of parameters such as the Hubble constant or
dark-energy that rely on large-scale homogeneity of the Universe.
It is however not easy to directly quantify the amount of scatter caused by peculiar velocities ({\it e.g.} see \cite{ali}).
On small  scales, {\it e.g.} scales of about  $10$ Mpc, the average deviation from the pure Hubble flow is precisely given by the variance of the peculiar velocities. If one were to constrain observations to
zones of minimum  peculiar velocities then one could assure a more precise determination of the global value of the Hubble parameter even from the ``very'' local measurements. Hence, the effect of sample variance can be hugely reduced if one could limit parameter determination to subsamples of velocity field where the variance of the peculiar velocities would be nearly vanishing. Here we will show, by using N-body simulations, that such sub-sampling can ensure that the uncertainty in the value of the Hubble parameter can be much smaller than if the whole sample were used.
This is in contrary to the usual approach which aims at obtaining a better statistic by averaging in a larger and larger sample indiscriminately.

The study of these minimum velocity dispersion zones, specifically their identification in large-scale surveys and their theoretical characteristics and classifications, are the subjects of this paper. On large scales, and even well into the nonlinear regime, the velocity field remains a potential flow and can be expressed as the gradient of a velocity potential \cite{arnold82,arnold84,NN}. In the linear Lagrangian regime and at high redshifts, where dark energy contribution to the expansion rate is small relative to the contribution from the dark matter, this is the same as the gravitational potential. Although the velocity field remains a potential flow well into the nonlinear regime and at low redshifts, the connection between the gravitational and the velocity potential becomes
more complex due to nonlinearities and also due to the influence of the dark energy on the evolution of the perturbations. Overall, and well into the nonlinear regime, the critical points of the gravitational potential yield the minimum velocity dispersion zones to an excellent approximation. These reside inside large-scale structures at:
the minima of the gravitational potential , {\it i.e.~} at the center of the over-densities such as galaxies and clusters or at the  maxima which are the low-density or void regions and finally at the
saddle points, which correspond to filaments and walls. However, our studies shows that zero-velocity zones are mostly to be found in the low-density or the void regions.

In the nonlinear regime, or in the multi-streaming regions, further zero-velocity zones can develop which would be more difficult to characterize theoretically, but can be figured out in N-body simulations.
We compare our theoretical predictions with a high resolution N-body simulation, with parameters taken from the Planck 2015 best $\Lambda CDM$ cosmological model. We pay a
special attention to the properties of the velocity field in linear and quasi-linear regimes in the vicinities of the voids on scales of about $5-10$ Mpc, where the existence of
the minimum velocity points with $|\textbf{v}|\rightarrow 0$ is more pronounced. We  show that these zones are the best
tracers of the Hubble flow, especially for determination of $H_0$ with accuracy of about a few percent.

The outline of the paper is as follows. In Section 2 we describe the mathematical basis of the method . In Section 3 we will briefly discuss the linear and non-linear evolution of velocity and density perturbations in the $\Lambda$CDM cosmological model . In Section 4 we introduce the concept of the minimum velocity zones as a basis for minimization of the uncertainties of the Hubble constant. In Section
5 we discuss an implementation of the method for N-body simulations, focusing on detection and estimation
of the variance of the Hubble flow in the vicinity of the critical points.  We shall show that these solutions are stable in time and can be easily traced back to the initial conditions. In Conclusion we summarize the results and discuss their  implementation for BAO.

 \begin{table}
\caption{The value of the Hubble parameter inferred from different observations}
\begin{tabular}{|l|l|}
\hline
Riess et al   2011 ( HST, cepheids) &    $73.8\pm 2.4$ \\
\hline
Freedman  et al 2012  (HST, Cepheids) &  $74.3 \pm 2.1$ \\
\hline
Reid et al  2013 (Masers) &  $68.9\pm 7.1$  \\
\hline
Hinshaw      2013  (9yr-WMAP)  &  $69.33 \pm 0.88$  \\
\hline
Efstathiou  2013 (Cepheids)  & $70.6 \pm 3 $\\
                                            & $72.5\pm 2.5$ \\
\hline
Planck collaboration  (2013,2015) &  $67.27\pm 0.66$ \\
\hline
Riess et al 2016 (WFC3 on HST, Cepheids) & $73.03 \pm 1.79$ \\
\hline
\end{tabular}
\end{table}

\section{Peculiar velocities and uncertainties of the Hubble flow.}
\label{sec:Peculiar velocities and uncertainties of the Hubble flow}
%-----------------------------------------------------------------

\begin{figure*}
%  \center
\includegraphics[width=0.45\textwidth]{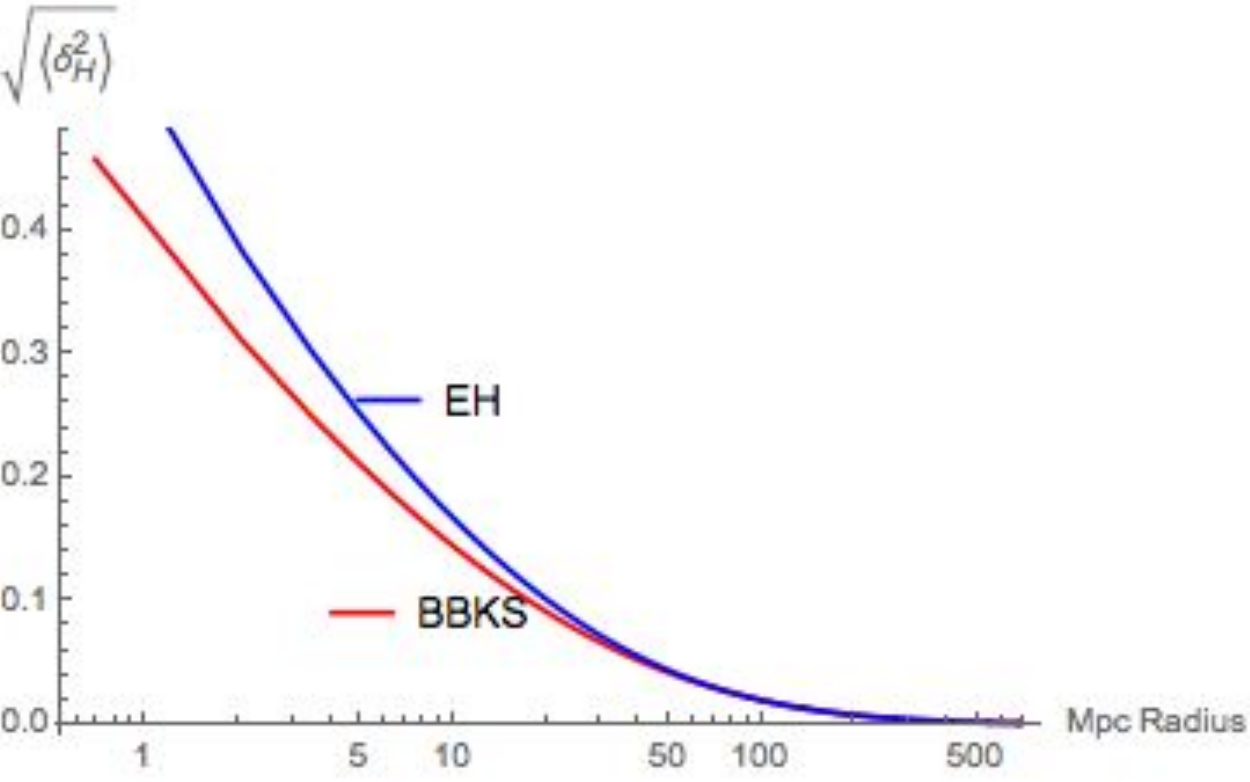}%}
%\centerline{
\includegraphics[width=0.45\textwidth]{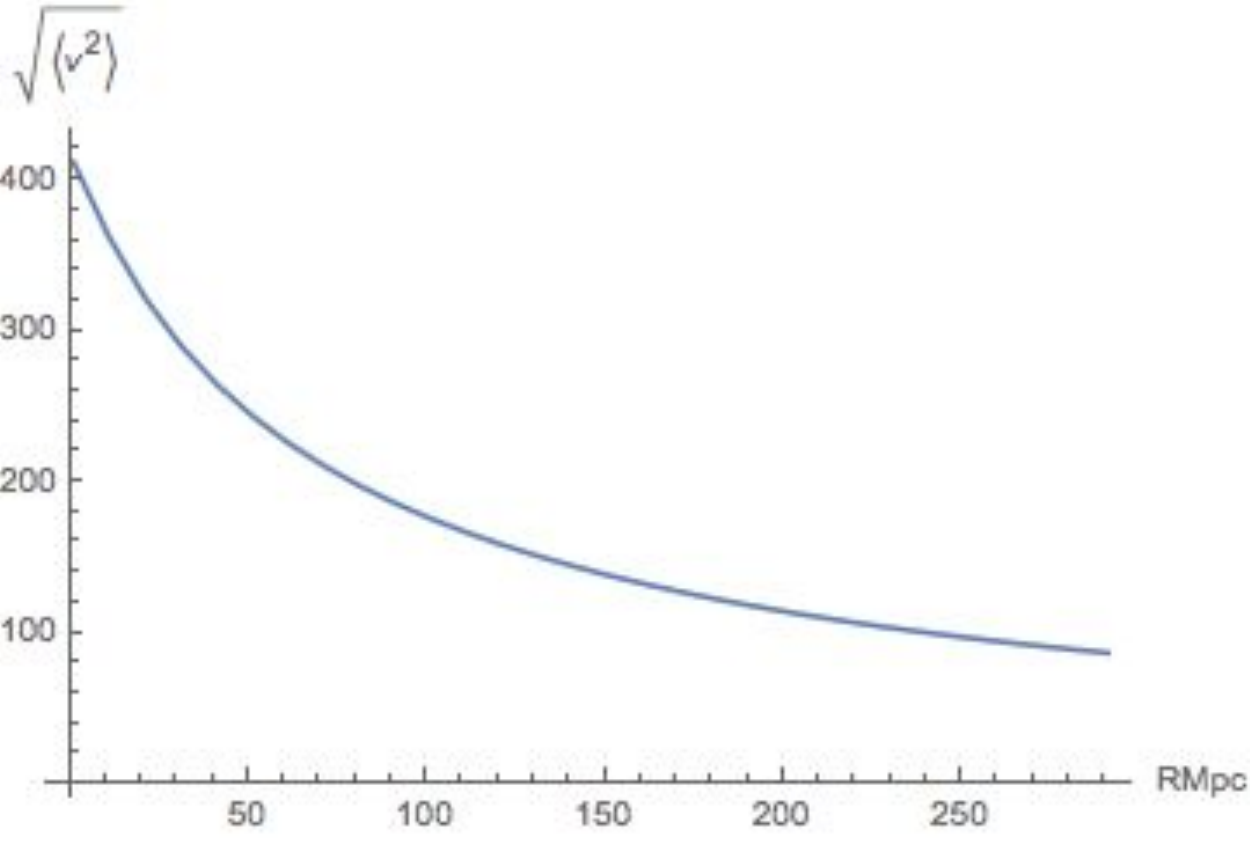}%}
\caption
{{\it left panel:} The variance of the Hubble flow as a function of radius is shown for two different types of transfer functions; BBKS and Hu and Eisenstein, the latter incorporates the
effect of baryons. {\it Right panel:} The variance of the velocity field as a function of the radius is plotted.}
\label{variance}
\end{figure*}

The Hubble law for an observer at position $\vec{r_j}=0$  can be expressed as
\beq
\vec{V}(\vec{r_i})=H(\vec{r_i})\vec{r_i}\,,
\label{eq1}
\eeq
where  the line-of-sight velocity is obtained from the redshift $z_i$ of the galaxy, $c\,z_i =\vec{V}_{||}(\vec{r_i})$        $\vec{r_i}=a(t)\vec{\chi}_i\quad$ with $\quad\vec{\chi}$ the comoving coordinates and  $a(t)$ the scalar factor.
 According to the Hubble flow,  at the same point we have :
\begin{eqnarray}
\vec{V}(\vec{r_i})=H_o\vec{r_i}+\vec{v}(\vec{r_i})\,,
\label{eq2}
\end{eqnarray}
where $\vec{v}(\vec{r_i})=a(t){\dot{\vec{\chi}}}_i$ is the peculiar velocity.

The comparison of eqn.~\ref{eq1} and eqn.~\ref{eq2} gives us the following estimator for the local fluctuation of the Hubble constant
\begin{eqnarray}
 \Delta(\vec{r_i})=\frac{H(\vec{r_i})-H_o}{H_o}=\frac{1}{H_o}\frac{\vec{v}(\vec{r}_i)\vec{r}_i}{|\vec{r}_i|^2}
\label{eq3}
\end{eqnarray}
For statistically isotropic and homogeneous random Gaussian field $\vec{v}(\vec{r_i})$ the function $\Delta(\vec{r_i})$ has
the same statistical properties with mean value $\langle\Delta(\vec{r_i})\rangle=0$ and the variance $\sigma^2=\langle\Delta^2(\vec{r_i})\rangle\neq 0$, where $<.>$ means an average over the statistical ensemble of realizations.
However, only one realization of the random field of the velocities is available to
us, which, due to structure formation in the Universe, is not
expected to satisfy a Gaussian distribution at all spatial scales. The separation of the
Gaussian and the non-Gaussian scales can be related to the correlation length usually taken as $r_c\simeq 5h^{-1}$Mpc, which is the scale at which
the two-point correlation function is unity,{\it i.e.} $\xi(r_c)=1$ \cite{peebles93}.  For $h=0.7$ this scale
is $r_c\simeq 7$ Mpc, and for scales larger than the correlation length, {\it i.e.} $r\gg r_c$, the density and velocity perturbations approximately evolve in the linear
regime. For $r\ll r_c $  the
density and velocity perturbations are non-linear and non-Gaussian. For all future analysis we will use the range of spatial
scale of about $5-10$ Mpc as a limit for the applicability of the quasilinear theories of the gravitational instability.
Thus, for $r\ge r_c$
 the subject of investigation is the mean value  of the random variable $\Delta_i=
\Delta(\vec{r_i})$ averaged over an sphere of radius $R>r_c$  and given by :
\begin{eqnarray}
 \Delta_R(\vec{x})=\int d^3y\frac{\vec{v}(y)\cdot(\vec{y}-\vec{x})}{H_o|\vec{y}-\vec{x}|^2}W(\vec{y}-\vec{x})\,,
\label{eq4}
\end{eqnarray}
where the window function $W(\vec{z})$ is given by the Heaviside function: $W(\vec{z})=\frac{3}{4\pi R^3}\Theta(R-|\vec{z}|)$ \citep{turner92,wang98} %\textcolor{red}{please put reference for she ?!}.
The velocity vector can be expressed in Fourier space since,
\begin{eqnarray}
 \vec{v}(\vec{y})=(2\pi)^{-3}\int d^3k \vec{v}_{\vec{k}}e^{-i\vec{k}\vec{y}}\,,
\label{eq5}
\end{eqnarray}
and the peculiar velocity $\vec{v}_{\vec{k}}$  and the density perturbations $\delta_{\vec{k}}$ in the linear regime are simply related through \cite{peebles93}:
\begin{eqnarray}
 \vec{v}_{\vec{k}}={f(\Omega)\,H_o\vec{k}\over {ik^2}}\delta_{\vec{k}}\,,
\label{eq6}
\end{eqnarray}
where $f(\Omega)=d{\rm ln} \delta/d{\rm ln}a\sim \Omega_0^{0.6} $ and we have taken bias of one on large scales and $\Omega_0$ is the present matter density normalized to the critical density.
The random variable $\delta_k$ is Gaussian in the linear regime. For $\Lambda CDM$ cosmological model the variance of $\Delta_R$ is given by the following relationship: $\sigma^2_H(R)\propto R^{-\gamma}$ , where $\gamma\simeq 2$ for $40h^{-1} \le R\le 100h^{-1}$ Mpc  \citep{wang98}. This result can be easily understood from the following estimations.
According to eqn.~\ref{eq3} we get:
\begin{eqnarray}
\sigma^2_H &\propto& \frac{3}{4\pi R^3H_o^2}\int d\phi d\theta\sin\theta\int drr^2\Delta^2(\vec{r_i})
 \nonumber\\
&\sim & \frac{3}{4\pi H^2_oR^3}\int d\phi d\theta\sin\theta\int_0^Rdr r^2\frac{(\vec{v}(\vec{r}) \vec{r})^2}{r^4}\simeq \frac{\sigma^2_v}{H^2_oR^2},\nonumber\\
\label{eq6a}
\end{eqnarray}
where $\sigma^2_v$ is the variance of the peculiar velocity in a sphere of radius R:
\begin{eqnarray}
 \sigma^2_v=\frac{1}{2\pi^2}\int d^3kP_v(k)
%\frac{3}{4\pi R^3}\int d^3r\prec v^2_i(\vec{r})\succ \Theta(R-|\vec{r}|)
\label{eq6b}
\end{eqnarray}
and $P_v(k)$ is the power spectrum of the peculiar velocities \cite{bbks}:
\be
&& P_v(k)=\overline{P}(k)T^2(k)
%&&  T(k)=\frac{\ln(1+aq)}{aq}\left[1+bq+(cq)^2+(dq)^3+(gq)^4\right]^{-\frac{1}{4}},\nonumber\\
\label{eq6c}
\end{eqnarray}
where $\overline{P}(k)$ is the primordial power spectrum, $T(k)$ is the tranfer function. %$a=2.34,~b=3.89,~c=16.1,~d=5.46,~g=6.71$,
with
%$\Gamma=\rho_r/1.68\rho_{cmb}$, $\Gamma=1$ for 3 flavors of relativistic neutrino plus CMB photons.
In Fig.~\ref{variance} the variance of the Hubble constant (left panel) and
the peculiar velocities (right panel) is plotted for
two different kinds of the transfer functions, BBKS  \citep{bbks} and Eisenstein and Hu \citep{eh98}.
At large radius the variance of the Hubble flow follows that of the peculiar velocities and becomes very small as is expected. Although, the effect on the variance is expected to be small even if there were
features in the power spectrum because of the smearing by the filter in the k-space~\citep{2010MNRAS.401..547H}.

\begin{figure*}
%  \center
\includegraphics[width=0.4\textwidth]{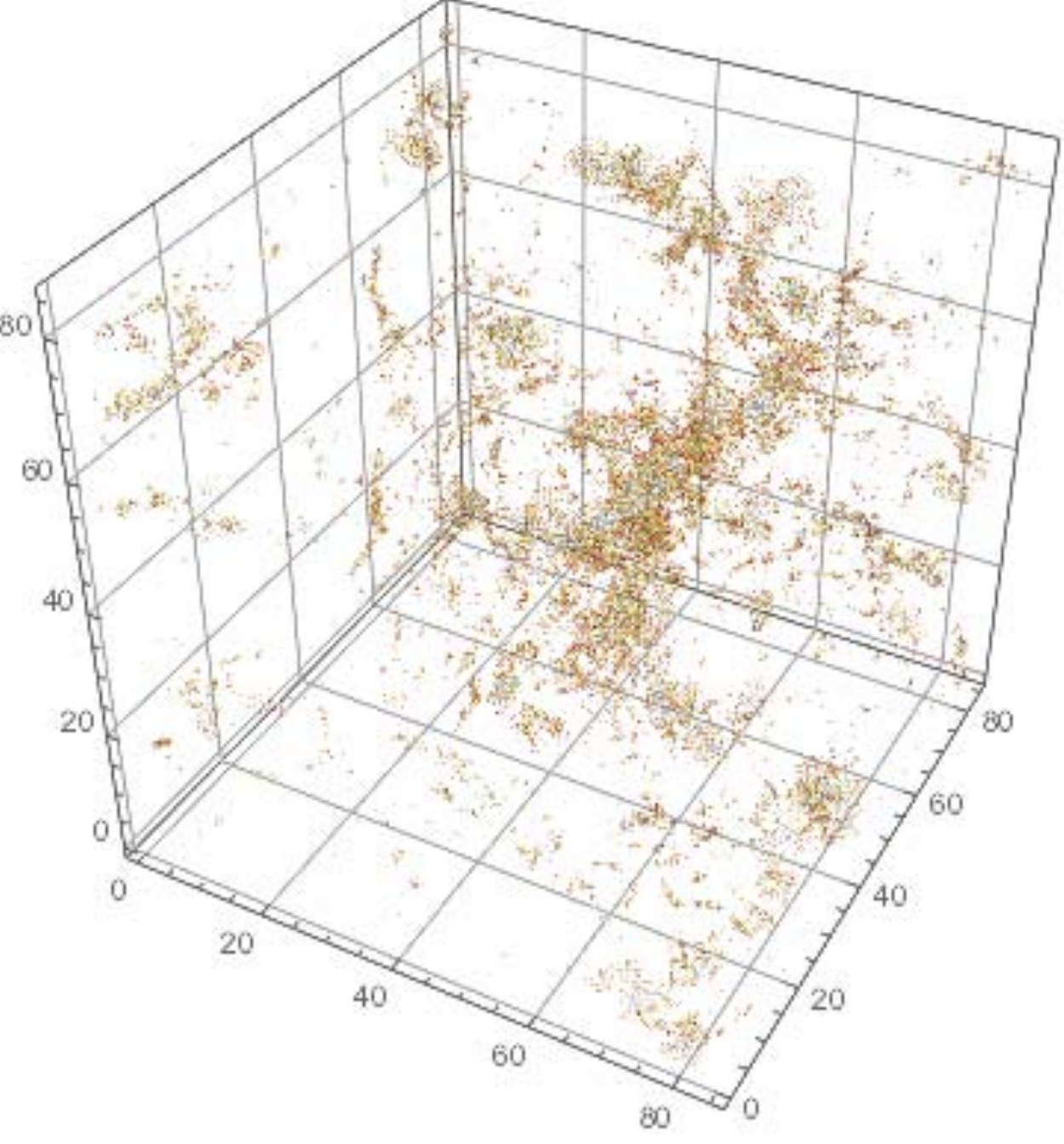}%}
%\centerline{
\includegraphics[width=0.65\textwidth]{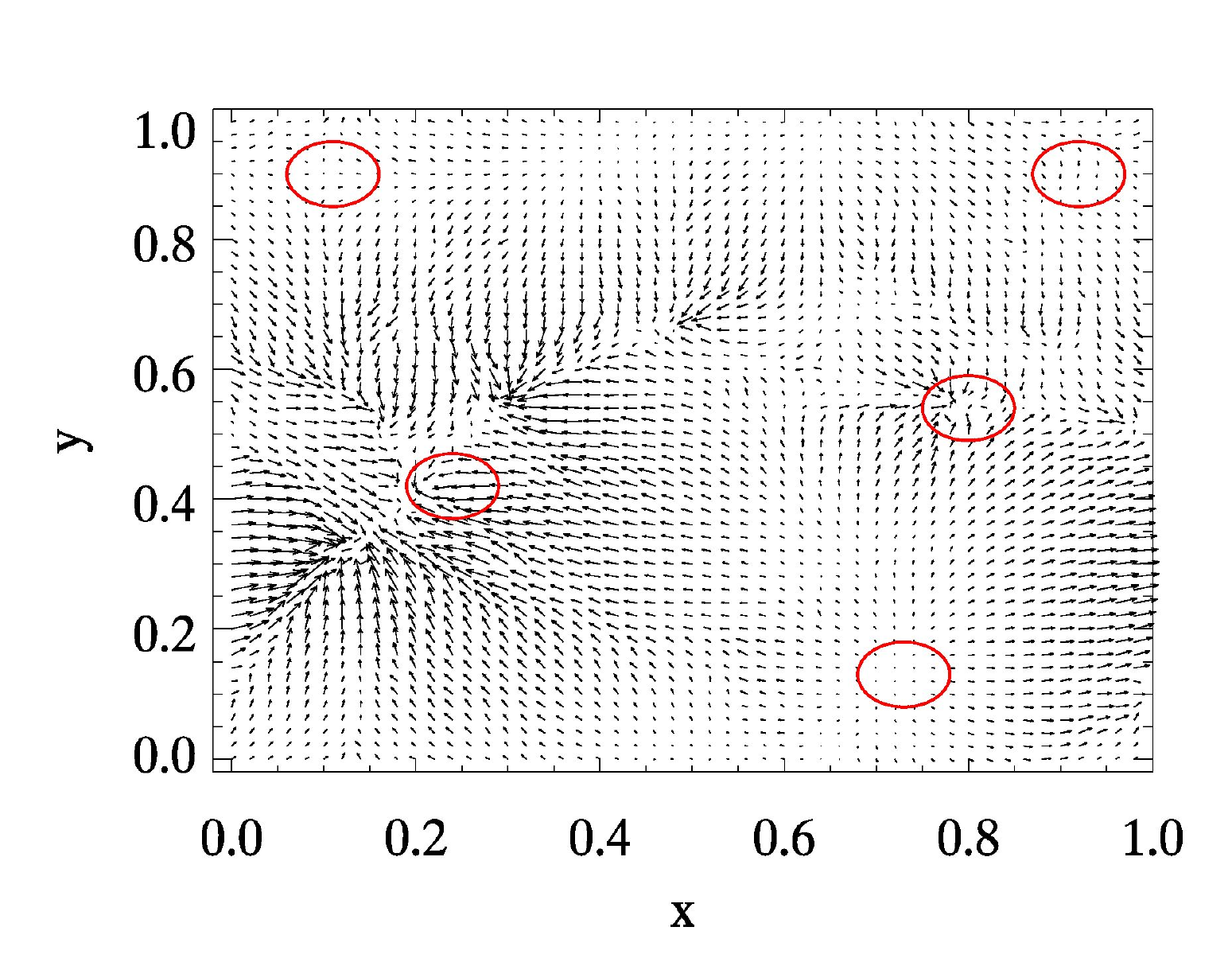}%}
\caption
{{\it Left panel} :
Distribution of the points of the velocity field with $|v|<50$km/s in the box of $85\times85\times 85$ Mpc
from N-body simulations at redshift $z=0$. {\it Right panel} : The 2D-slice  along
the plane at the hight $Z=0.5\pm 0.1$ with different critical points of the velocity marked by the red circles. The X,Y axes are normalized to
85 Mpc.
}
\label{fig0}
\end{figure*}

As it is seen from  eqn.~\ref{eq6a}, the variance $\sigma^2_H$ is
 the ratio between $\sigma^2_v$
and the square of the Hubble velocity at  distance $R$. This is why
the estimator (\ref{eq6a}) shows clearly that an efficient way to reduce the uncertainties for $\sigma^2_H$ in a given volume, is by taking a subsample of specific zones where
the amplitudes of the velocities are small. Furthermore, as we shall show the morphology of these zones is extremely stable in time and hence the memory of the initial condition is well kept in these regions of vanishing velocities. Next, we try to quantify these zones of minimum peculiar velocities using the standard first-order Lagrangian perturbation theory.

\section{The velocity and the density fields in the Zel'dovich approximations}
%-------------------------------------------------------------------

The zones of minimum velocity on large-scales can be classified using first-order Lagrangian perturbation theory which stays valid well into the nonlinear regime, {\it i.e.} into $\delta\rho/ \rho > 1$. However, in the nonlinear regime further zones of vanishing velocities emerge which are due to multi-streaming or shell-crossing. The work presented here is meant for large-scales and hence does not classify zones of vanishing velocities in the multi-stream regions.
 The  fluid equation relating the velocity $v_i$, density $\rho$ and
perturbations of the gravitational potential $\phi$ are \citep{peebles80,shandarin89} \,,
\begin{eqnarray}
&&\frac{\partial v_i}{\partial t}+\frac{1}{a}(v_k\frac{\partial}{\partial\chi_k})v_i+H(t)v_i=-\frac{1}{a}\frac{\partial\phi}{\partial\chi_i},\nonumber\\
&&\frac{\partial^2\phi}{\partial\chi^2_i}=4\pi Ga^2[\rho(\chi_i,t)-\overline\rho(t)]=4\pi Ga^2\overline\rho(t)\delta(\chi_i,t)
,\nonumber\\
&&\frac{\partial\rho(\chi_i,t)}{\partial t}+3H(t)\rho(\chi_i,t)+\frac{1}{a}\frac{\partial}{\partial\chi_i}
(\rho(\chi_i,t)v_i)=0\,,\nonumber\\
\label{eqS}
\end{eqnarray}
where $H(t)=\frac{\dot{a}}{a}$, $\bigtriangledown_i=\frac{\partial}{\partial\chi_i}$, and $v_i=a(t)\dot{\chi}_i=a(t)U_i$.
In the linear approximation in the CDM-dominated epoch  $\delta\propto a(t)$, and  from linearization of the fluid equations one can easily deduce that the gravitational potential is time-independent in the linear regime {\it i.e.} $\bigtriangleup\phi=F(\vec{\chi})$.  The Euler equation  in the linear regime yields
\begin{eqnarray}
 v_i(t,\vec{\chi})\simeq -\frac{t}{a(t)}\frac{\partial\phi}{\partial\chi_i}=a(t)\dot{\chi}_i\,,
\label{eqV}
\end{eqnarray}
whose solution, found iteratively, is
\begin{eqnarray}
 \chi_i(t)\simeq  q_i-\eta(t)\bigtriangledown\phi(q_i)
\label{eqI}
\end{eqnarray}
where: $q_i=\textbf{q}$ and $\bigtriangledown\phi(q_i)$ are the coordinates $\chi_i$ and the gradient $\frac{\partial\phi}{\partial\chi_i}$ at $t=0$, and $\eta(t)=\int dt\,t/a^2(t)$.
Expresion (\ref{eqI}) is the well-known Zel'dovich approximation which provides a solid theoretical framework for
the understanding of the evolution of the density and velocity fields into the non-linear regime \cite{zeldovich70}.
In the usual form  the Zel'dovich approximation is given by:
\begin{eqnarray}
 \textbf{r}(t,\textbf{q})= a(t)\left(\textbf{q}+\eta(t)\nabla\Phi(\textbf{q})\right)\,,
\label{eq7}
\end{eqnarray}
where $\textbf{q}$  is the Lagrangian coordinate, $\textbf{r}(t,\textbf{q})$ is the physical Eulerian coordinate of the particles at the time t,  $\Phi(\textbf{q})=-\phi(\textbf{q})$ is the velocity potential at $t=0$:
\begin{eqnarray}
v_i=a(t)\dot{\eta}\bigtriangledown_i \Phi(\textbf{q})
\label{eqZA}
\end{eqnarray}

Thus in the Zel'dovich approximation the morphology of the velocity field directly reflects the initial ($t=0$) spatial distribution of the gradient
$ \bigtriangledown_i\Phi(\textbf{q})$, with the amplitude simply evolves as $ a\dot{\eta}$.

From eqn.~\ref{eq7} and eqn.~\ref{eqZA} clearly seen that the points of extreme $\bigtriangledown_i \Phi(\textbf{q})=0$ are peculiar, since
in the vicinity of these points $\Delta q_j $ the  velocity potential can be Taylor expanded up to the third order of magnitude as:

\begin{eqnarray}
 \Phi(q_i)\simeq \Phi_0(q_i^{ext})+\frac{1}{2}\sum_{j,k}\alpha_{j,k}\Delta q_j\Delta q_k
+\frac{1}{3!}\sum_{j,k,n}\beta_{j,k,n}\Delta q_j\Delta q_k\Delta q_n
\label{eqEX}
\end{eqnarray}
where $\Delta q_j=q_j-q_j^{ext}$, $\alpha_{j,k}=D_{j,k}(q_j^{ext})=\frac{\partial \Phi}{\partial q_j\partial q_k}$ is the deformation tensor,
$\beta_{j,k,n}=\frac{\partial^3\Phi}{\partial{\Delta q_j} \partial{\Delta q_k}\partial{\Delta q_n}}$  is a matrix of the coefficients of the third  order expansion at
$\Delta q_k=0$,  $\Phi_0(q_i^{ext})$ are  constants, which can be set to zero without lost of generality.

Taking eqn.~\ref{eqEX} under consideration, lets discuss  two different asymptotic of the potential $\Phi(\textbf{q})$ in the vicinity of $q_j^{ext}$.\\

\textbf{ 1. Non-vanishing eigenvalues of the deformation tensor.}\\
Suppose, that the eigenvalues of the deformation tensor are non vanishing, and  $|\Delta q_j|$ is small in comparison to the correlation length
$L_c$ for the velocity field at the linear regime. In this case we can neglect of the third term in eqn.~\ref{eqEX} and get the following representation for the velocity field around the critical points:
\begin{eqnarray}
v_i\simeq a(t)\dot{\eta}\sum_k\alpha_{i,k}\Delta q_k,\hspace{0.2cm} \chi_i(t)\simeq  q_i+\eta(t)\sum_k\alpha_{i,k}\Delta q_k
\label{ez}
\end{eqnarray}
%and $\chi_i\simeq q_i$ .
%and, correspondingly:
%\begin{eqnarray}
 %\chi_i(t)\simeq  q_i+\eta(t)\sum_k\alpha_{i,k}\Delta q_k
 %\label{eqZA2}
%\end{eqnarray}

The density field of the cold dark matter, $\rho_{cdm}$, in the Zel'dovich approximation which simply follows from the continuity or the conservation of mass equation, is  given by
\begin{eqnarray}
 \rho_{cdm}(\chi_j,t)={\rho(t)\over \left |\delta_{ik}+\eta(t)\alpha_{i,k}\right |},
\label{eq8}
\end{eqnarray}
where $\rho(\chi_j,t)\simeq  \rho(t)$ is the  density at $\eta=0$ at the critical point, where  $\Delta q_j=0$. %, and
 %$D_{ik}=\frac{\partial^2\Phi(\textbf{q})}{\partial\textbf{q}_i\partial\textbf{q}_k}$ is the deformation tensor of the velocity potential.
Using the linear transformation of the coordinates in the vicinity of the critical point, $q_i\rightarrow \varepsilon_i$, the matrix $\alpha_{i,k}$
can be diagonalize as : $\alpha_{i,k}=\lambda_i\delta_{i,k}$, where $\delta_{i,k}$ is the Kronecer symbol. Then, from eqn.~\ref{ez} and eqn.~\ref{eq8} we get:

\begin{eqnarray}
v_i\simeq a(t)\dot{\eta}\lambda_i\varepsilon_i, \hspace{0.2cm} \rho_{cdm}(\varepsilon_i,t)=\frac{\rho(t)}{ \Pi_i(1+\eta(t)\lambda_i)}
\label{ez1}
\end{eqnarray}
where $\Pi_i x_i\equiv x_1x_2x_3$.
Zel'dovich approximation has been mainly and extensively used in the past to characterize the cosmic density field. This approximation is valid for as long as the determinant in the denominator of  eqn.~\ref{ez1} has not become zero, {\it i.e.} in the single-stream regions. The denominator can vanish when one, two or all three eigenvalues $\lambda_i$ become negative, and $\Pi_i\lambda_i<0$.

{\bf a.} Along the axis with
$|\lambda_1|\gg |\lambda_2|,|\lambda_3| $ with negative eigenvalue of $\lambda_1<0$ the denominator in eqn.~\ref{ez1} will vanish first , forming the so-called Zel'dovich pancakes or sheets. In this case the velocity field is characterized by eqn.~\ref{ez1} with the major component
$|v_1\simeq a(t)\dot{\eta}\lambda_i\varepsilon_1|\gg |v_2|,|v_3|$ and $v_1<0$. At the moment $t_{1}$, when $1+\eta(t_1)\lambda_1=0$,
the corresponding velocity fiels is very asymmetric , and oriented along the $\varepsilon_1$- axis towards the critical point. The amplitude
of the $v_1$-component at $t=t_1$ is given by:
\begin{eqnarray}
v_1\simeq -\omega(t_1)\varepsilon_1,\hspace{0.2cm}v_2=-\omega(t_1)\frac{\lambda_2}{\lambda_1}\varepsilon_2,
\hspace{0.2cm}v_3=-\omega(t_1)\frac{\lambda_3}{\lambda_1}\varepsilon_3
\label{v1}
\end{eqnarray}
where $\omega(t_1)=a(t_1)\frac{\dot{\eta}(t_1)}{\eta(t_1)}$.  \\

{\bf b.} Once the second denominator vanishes ($\lambda_2<0, |\lambda_2|\gg |\lambda_3|$ ) then filaments form due to collapse along two axes and if the gravitational collapse proceeds along all three axes then nodes would form.
The all negative $\lambda_i<0$ corresponds to the point of maxima of the velocity potential, and correspondingly, the point of minima for
the gravitational potential at early stages of gravitational instability.  Note, that the existence of the anisotropic flow is an important feature of the quasilinear approximation, based on  non-linear Zeldovich approach for the density and  linear description of the
velocity field around critical points.

{\bf c.}  The case when $\Pi_i\lambda_i>0$ corresponds to completely different properties of the density and velocity fields in the quasilinear regime.
All positive $\lambda_i>0$ guaranty an absence of singularities for the density fields, as it is seen from eqn.~\ref{ez1}, and the corresponding critical points are associated with the points of minima of the velocity potential $\Phi(\textbf{q})$ (maxima for the gravitational potential $\phi$).
The non-linear stages of evolution of the density field will form a voids, where $\rho_{cdm}\simeq \rho(t)/(\eta^3\Pi_i\lambda_i)\ll \rho(t)$, and
the direction of the  velocity  flow , given by eqn.~\ref{ez1}, is pointed outwards from the critical points.
\\
      {\bf 2. Vanishing eigenvalues of the deformation tensor.}

In the case of degeneracy of the deformation tensor, the velocity flow is given by the cubic term in eqn.~\ref{eqEX}, which corresponds to the
saddle point of the velocity potential $\Phi$ and the gravitational potential $\phi$. For a saddle points the corresponding velocity flow is given by the quadratic form:
\begin{eqnarray}
v_i\simeq \frac{a(t)\dot{\eta}}{3!}\sum_{j,k}\left[\beta_{i,j,k}+\beta_{j,i,k}+\beta_{j,k,i}\right]\Delta q_j\Delta q_k
\label{saddle}
\end{eqnarray}
In depends on the properties of the $\beta_{i,j,k}$ coefficients, the quadratic form can be elliptic or hyperbolic, which determines very complex structure of the velocity and density fields. However, as it was shown in~\cite{dorr}, the probability for existence of the saddle
points of the potential is in order of factor $(\Pi_i\lambda_i)^2\rightarrow 0$, if $\lambda_i\rightarrow 0$. This is why in our future analysis we will focus on the properties of the voids, where the velocity flow is characterized by   relatively low amplitudes , and consequently, lower
uncertainties  of the Hubble parameter $H_0$.

\section{The variance $\sigma_H^2$ in the vicinity of the critical points: analytical approach.}
%--------------------------------
The properties of the  velocity field  in the vicinity
of the critical points, associated with voids,  allows us to estimate the variance  $\sigma_H^2$ . As we have shown in Section 1, $\sigma_H^2$ is given by  eqn.~\ref{eq3} and  eqn.~\ref{eq6a}. When studying the stationary points associated with a void, we need to take under consideration that in the reference system  with the origin at this stationary point (see Fig.~\ref{figR}), the distribution of the velocities is similar to an anisotropic Hubble flow  (see eqn.~\ref{ez1}):
\begin{figure}[!htb]
%  \begin{center}
\hbox{
    \centerline{\includegraphics[scale=0.3]{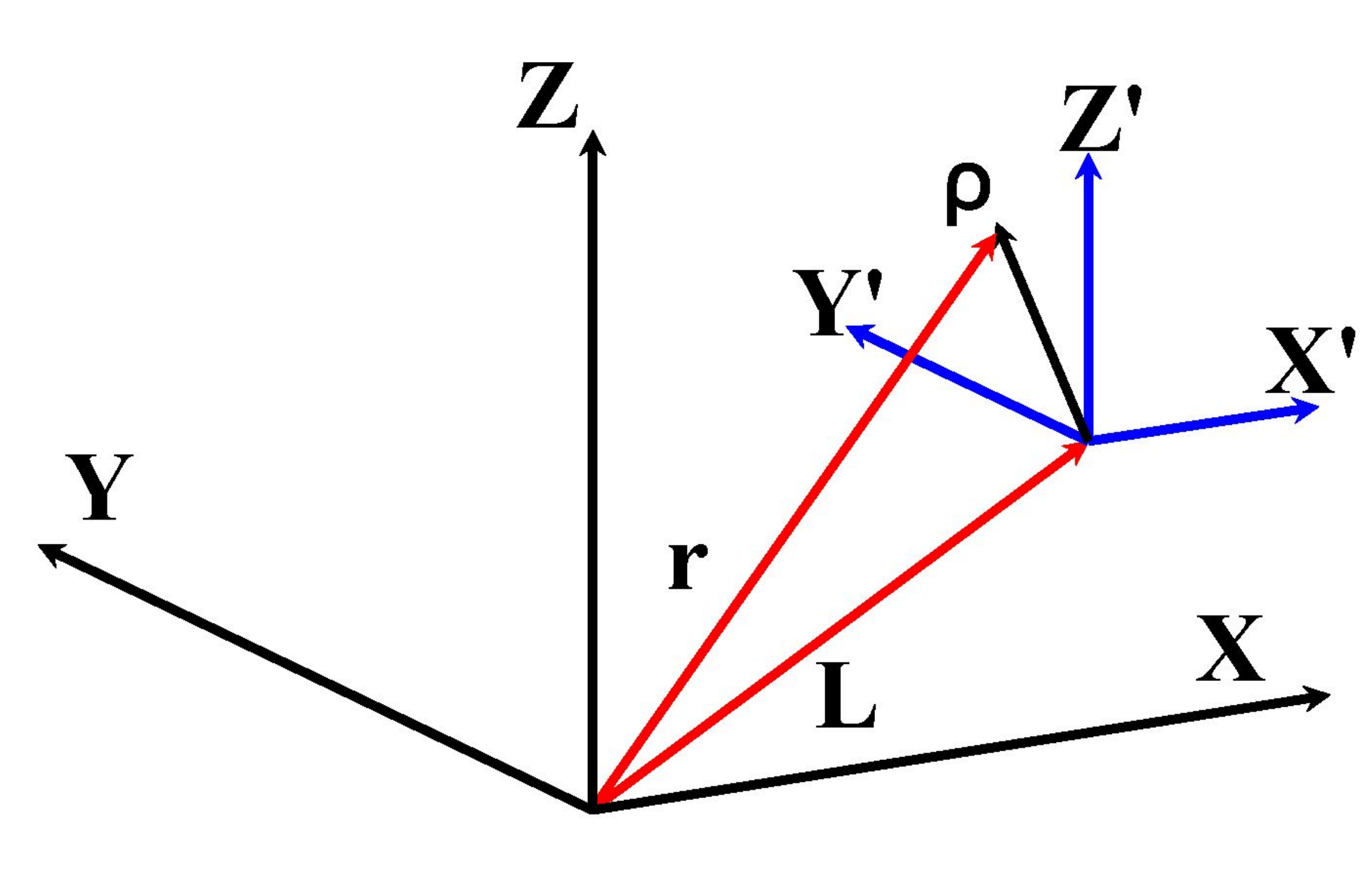}}}
    \caption{The reference system of coordinates $X,Y,Z$ associated with the position of an observer  and the system $X',Y',Z'$, with origin the
    at the critical point of the velocity flow. }
    \label{figR}
%  \end{center}
\end{figure}

\begin{eqnarray}
 V_i=\xi_i{\cal L}_i
\label{nbod}
\end{eqnarray}
where : $\xi_i=a(t)\dot{\eta}\lambda_i$, ${\cal L}_i=\varepsilon_i$, ${\cal L}_1={\cal L}\sin\theta\cos\phi$, ${\cal L}_2={\cal L}\sin\theta\sin\phi$, ${\cal L}_3={\cal L}\cos\theta$ are the
coordinates of the point inside the spherical zone, centered around an stationary point.
% As it is seen from  eqn.~(\ref{nbod}), there is an anisotropic regular ``Hubble'' expansion  for an observer
%located at the stationary point.
 Assuming that the position
of the center of the stationary point in the reference frame of the observer at $\vec{r}_j=0$ is given by the vector
$\vec{l}_i$, and $\vec{r_i}=\vec{l}_i +\vec{\cal L}_i$, we can estimate $\Delta(\vec{r}_i)$ as follows:
\begin{eqnarray}
&& \Delta(\vec{\cal L}_i)=\frac{1}{H_0}\frac{V_i\vec{r}_i}{|\vec{r}_i|^2}\simeq \frac{V_ie_i}{H_0L}%+\nonumber\\
%&&
+\frac{V_i{\cal L}_i}{H_0L^2}-\frac{2}{H_0L^2}(V_ie_i)({\cal L}_je_j),\nonumber\\
\label{vec}
\end{eqnarray}
where $L=|\vec{r}_i|$, $e_i=\vec{l}_i/|\vec{l}_i|$, and in the scalar product we have used the following
definition:$V_ie_i\equiv \sum_i(V_ie_i)$.
For the future analysis we will make an average over the sphere with radius $\varrho$ and the corresponding
volume $4\pi\varrho^3/3$:
\begin{eqnarray}
 \prec\Delta(\vec{\cal L}_i)\succ &=& \frac{3}{4\pi{\varrho}^3}\int_0^{\varrho}d{\cal L}{\cal L}^2\int_0^{2\pi}d\phi\int_0^{\pi}d\theta\sin\theta \Delta(\vec{\cal L}_i)\nonumber\\
&\simeq & \frac{{\varrho}^2}{5H_0L^2}\left[\xi_x(1-2e^2_x)+\xi_y(1-2e^2_y)+\xi_z(1-2e^2_z)\right]\,.\nonumber\\
\label{vec1}
\end{eqnarray}
In the case when $\xi_x\simeq \xi_y\simeq \xi_z=\xi$ we have
\begin{eqnarray}
 \Delta_{\varrho}=\prec\Delta(\vec{\cal L}_i)\succ\simeq -\frac{3\xi{\varrho}^2}{5H_0L^2}\simeq -\frac{\varrho}{L}\frac{V_{\varrho}}{V_H}\,,
\label{vec2}
\end{eqnarray}
where $V_{\varrho}$ is the peculiar velocity at the edge of the sphere with radius $\varrho$, and $V_H$ is the
Hubble velocity at the distance L from the observer at $\vec{r}_j=0$.

From  eqn.~\ref{vec} we can get the variance for the $\Delta(\vec{\cal L}_i)$-function averaged over the sphere with the radius $\varrho$ as :
\begin{eqnarray}
 \sigma^2_{\varrho}=\prec\left[\frac{(V_ie_i)^2}{H_0^2L^2} \right]\succ-(\prec\Delta(\vec{\cal L}_i)\succ)^2\simeq \nonumber\\
\simeq \frac{\varrho^2}{5H_0^2L^2}(\xi^2_xe^2_x+\xi^2_ye^2_y +\xi^2_ze^2_z)\simeq\frac{V^2_{\varrho}}{5V_H^2}\ll 1\,,
\label{vec3}
\end{eqnarray}
for $\xi_x\simeq \xi_y\simeq \xi_z$.
From the comparison of  eqn.~\ref{vec3} and  eqn.~\ref{eq6a} we get: $\sigma^2_{\varrho}/\sigma^2_H\propto V^2_{\varrho}/\sigma^2_v\ll 1$. Thus, the estimation of the uncertainties of  the Hubble
constant at the critical points, localized inside the sphere with radius $R=L$, will provide smaller
rms  then for the whole volume.

%\\
 %\textcolor{red}{-----------------------------------------------------------------------------------------------------------------}
\section{Critical points of the velocity potential in the $\Lambda$CDM model: N-body simulations}

\begin{figure}
\centerline{
\includegraphics[width=0.43\textwidth]{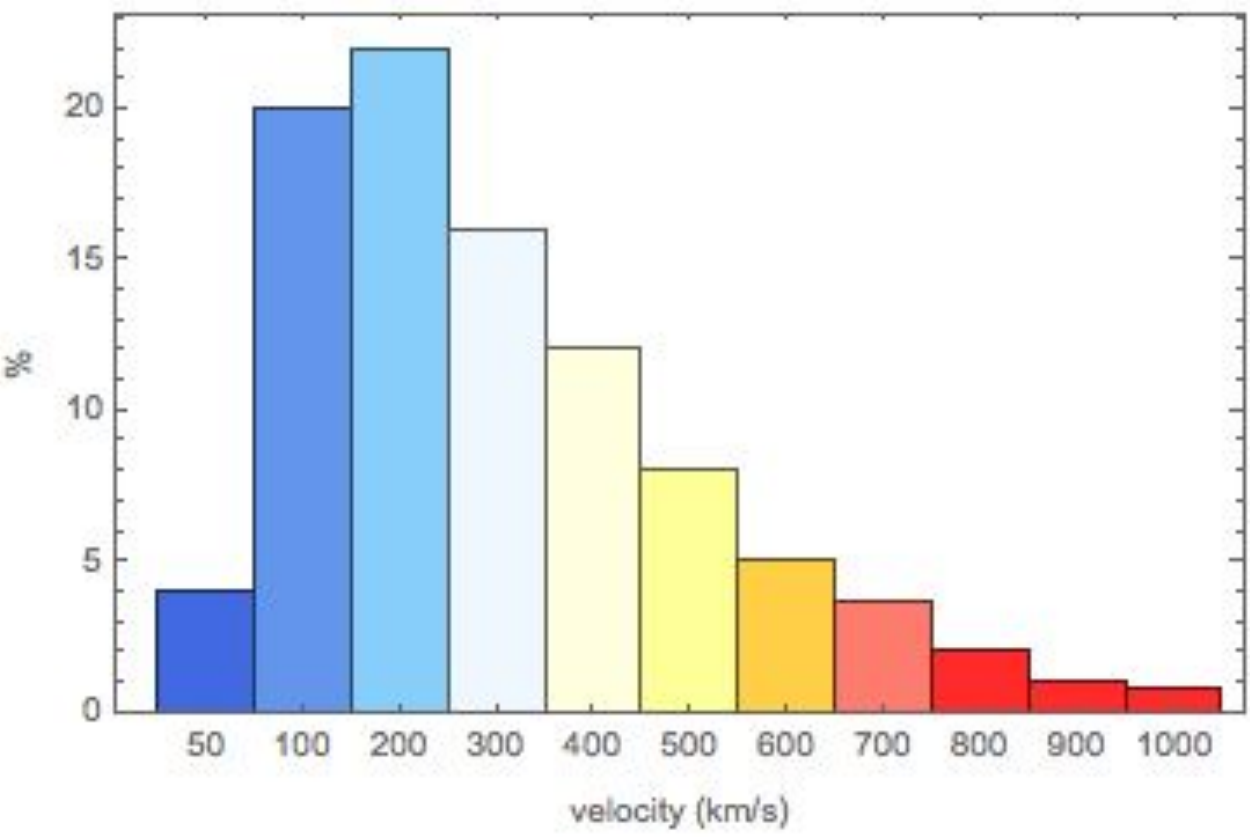} %{velocitiies-histogram-percentage-slicez-thinner.pdf}
\includegraphics[width=0.28\textwidth] {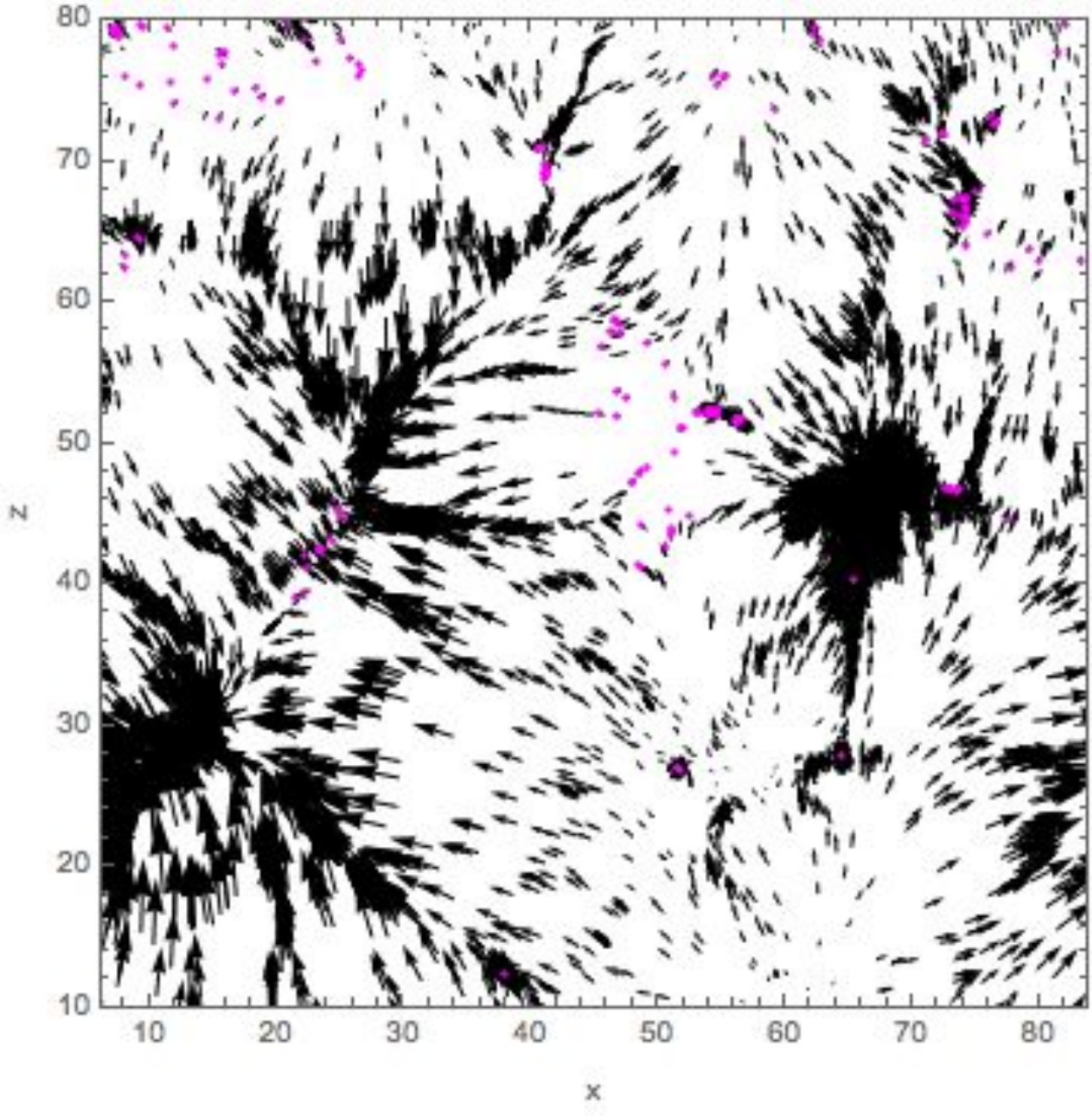}
\includegraphics[width=0.28\textwidth] {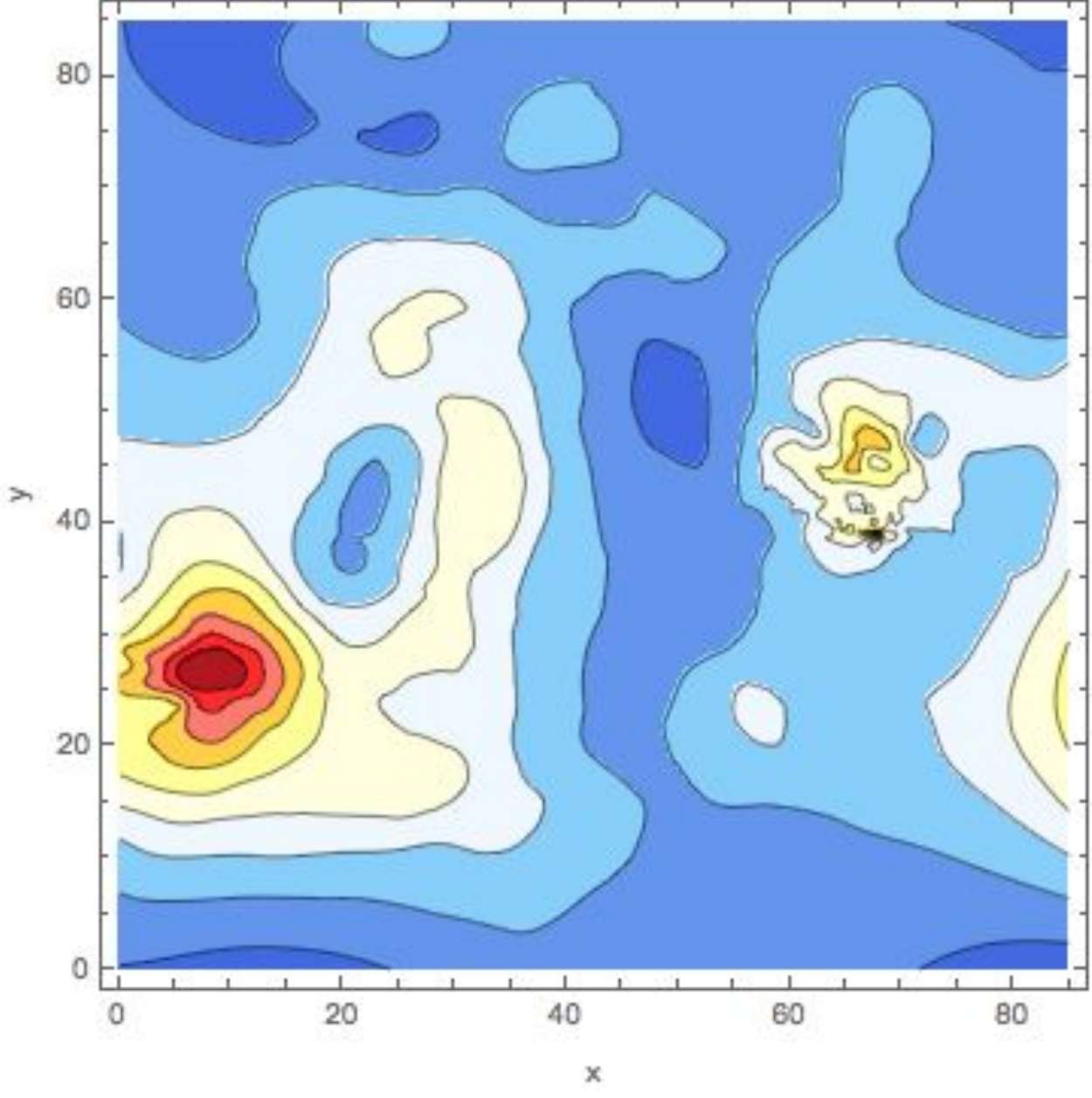}}
\caption{{\it left panel:} Histogram of the velocities taken from the simulation box with $256^3$ particles at the redshift z=0. About four percent of the particles are in very cold regions of velocities less than 50 km/s.
{\it Middle panel:} 2D sampling of the velocity distribution  in a thin slice $Z_c=0.4125\pm 0.0025$ of the simulation box, with vectors in magenta representing those with velocities less than 50 km/s. {\it Right panel:} A contour plot of cold regions presented.
% from the full simulation box.
%
%a thin slice of the simulation box
%plots are the same as those presented in \ref{histo-contor-part-full} with the difference that here all measurements are made in a thin slice of the simulation box and the contour plot is made for the full range of velocities and its color legend given by the histogram.
}
\label{histogram-velocity}
\end{figure}
\begin{figure*}[!htb]
%\begin{center}
\centerline{
\includegraphics[width=0.32\textwidth]{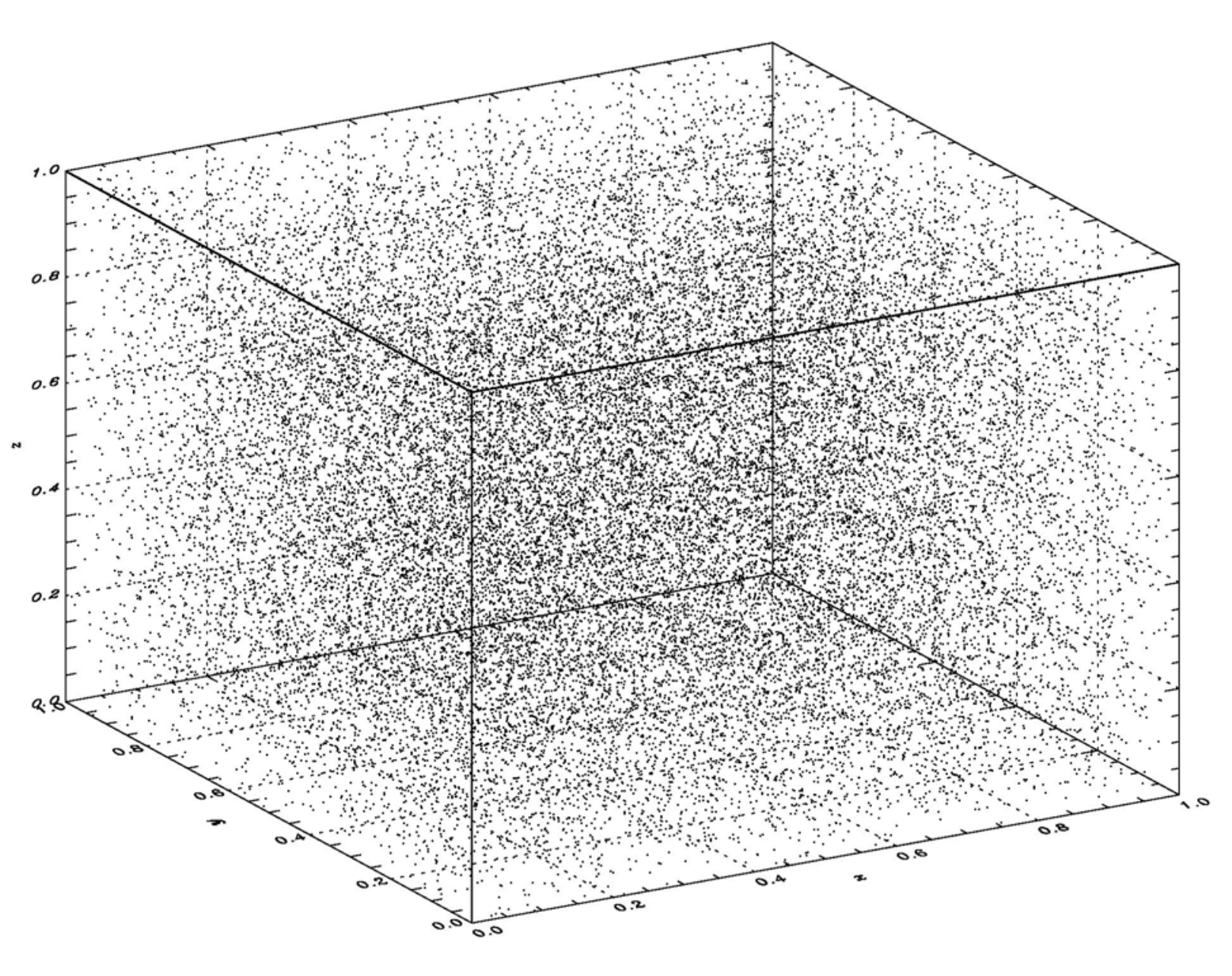}
\includegraphics[width=0.32\textwidth]{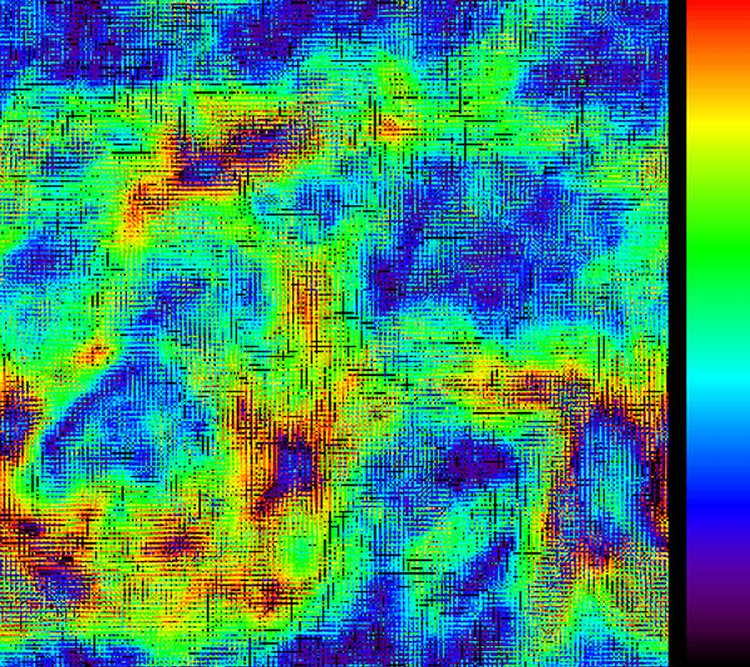}
\includegraphics[width=0.32\textwidth]{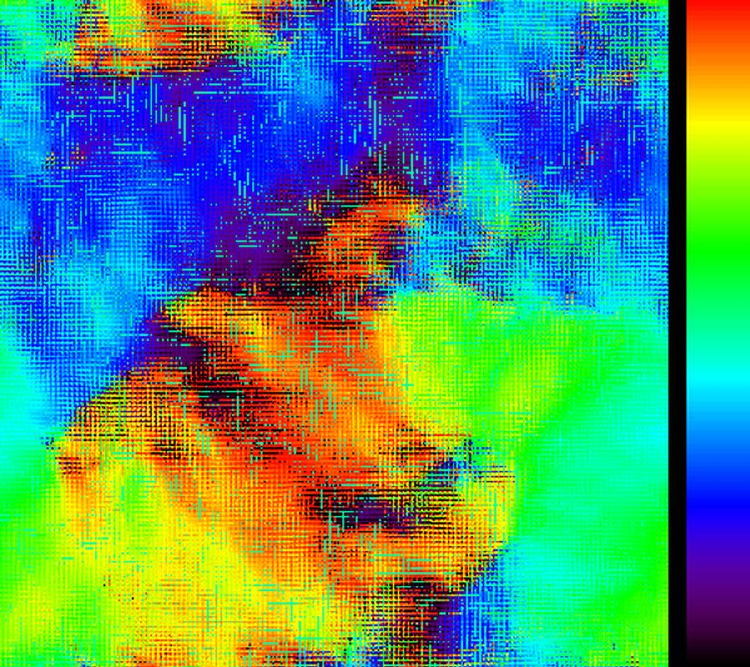}}%}
\centerline{
\includegraphics[width=0.32\textwidth]{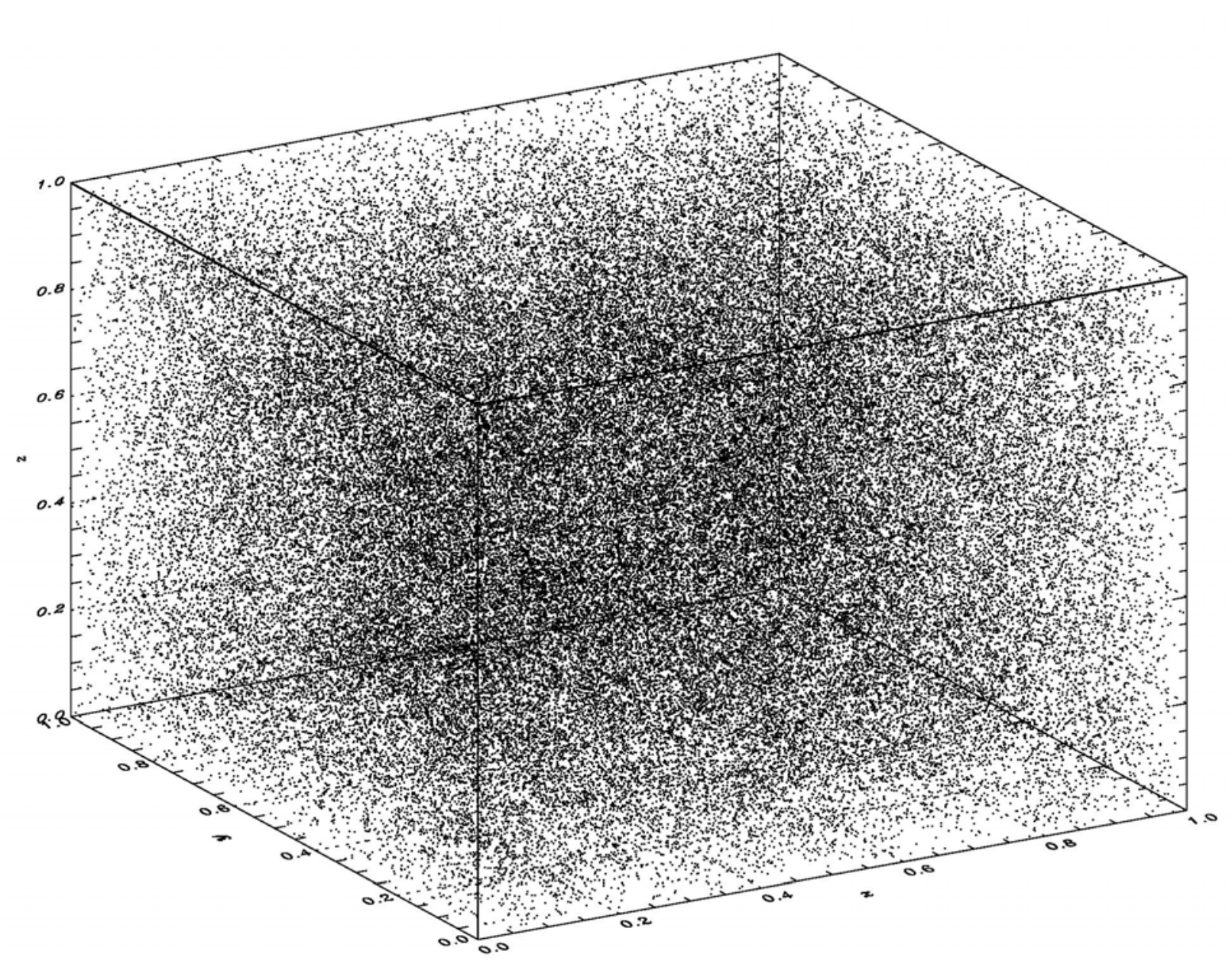}
\includegraphics[width=0.32\textwidth]{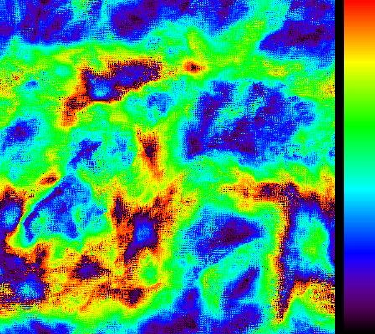}
\includegraphics[width=0.32\textwidth]{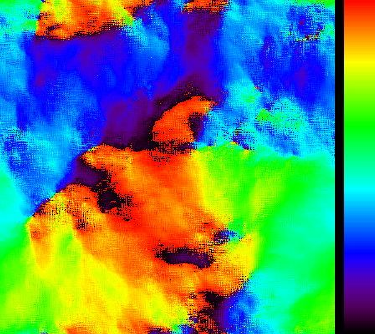}}%}
\centerline{
\includegraphics[width=0.32\textwidth]{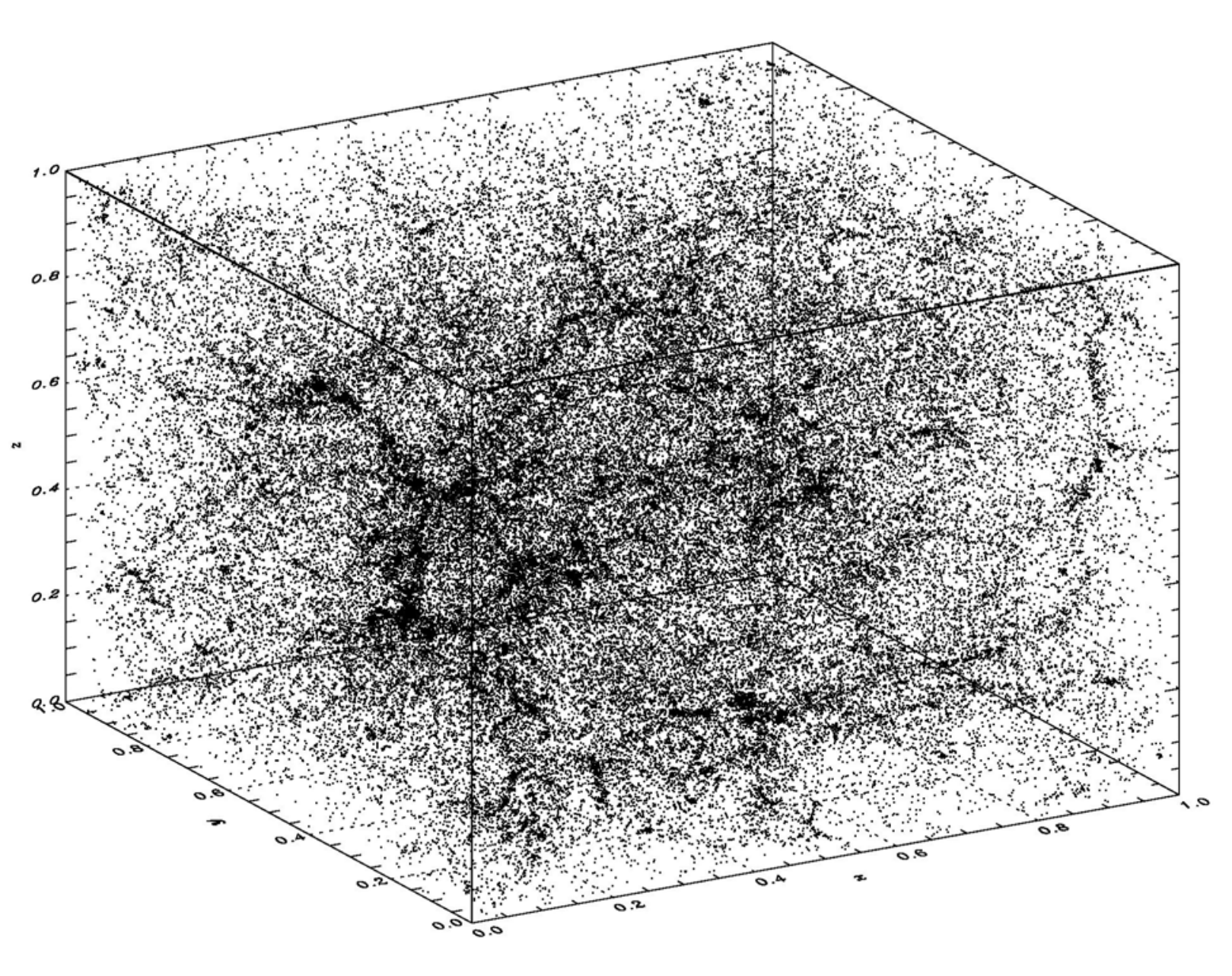}
\includegraphics[width=0.32\textwidth]{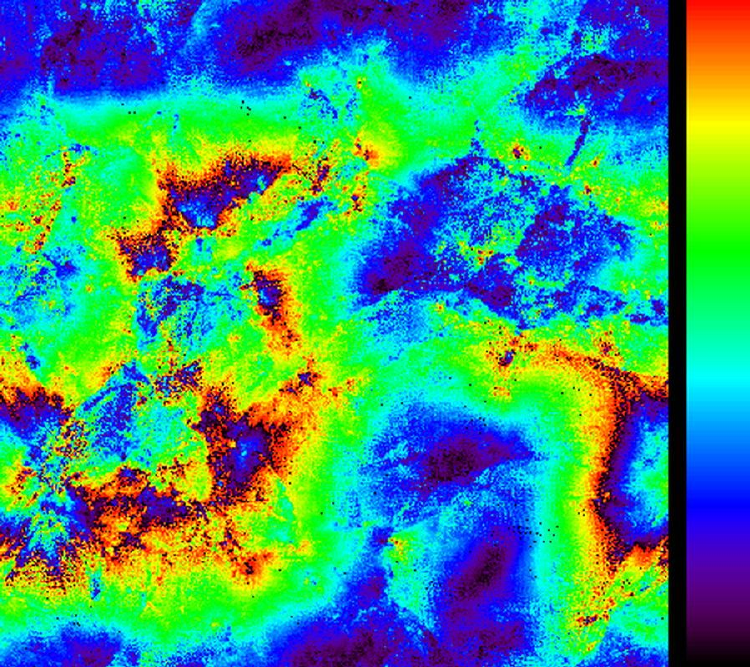}
\includegraphics[width=0.32\textwidth]{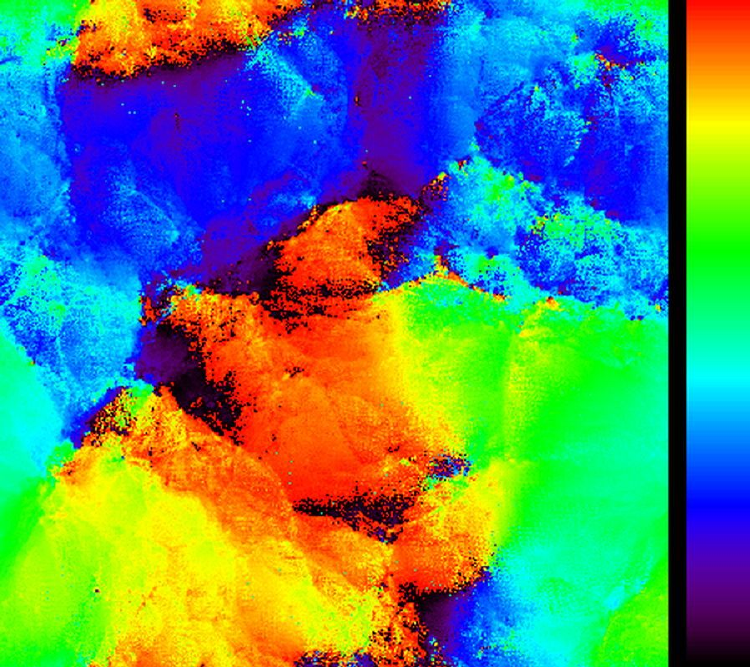}}%}
\centerline{
\includegraphics[width=0.32\textwidth]{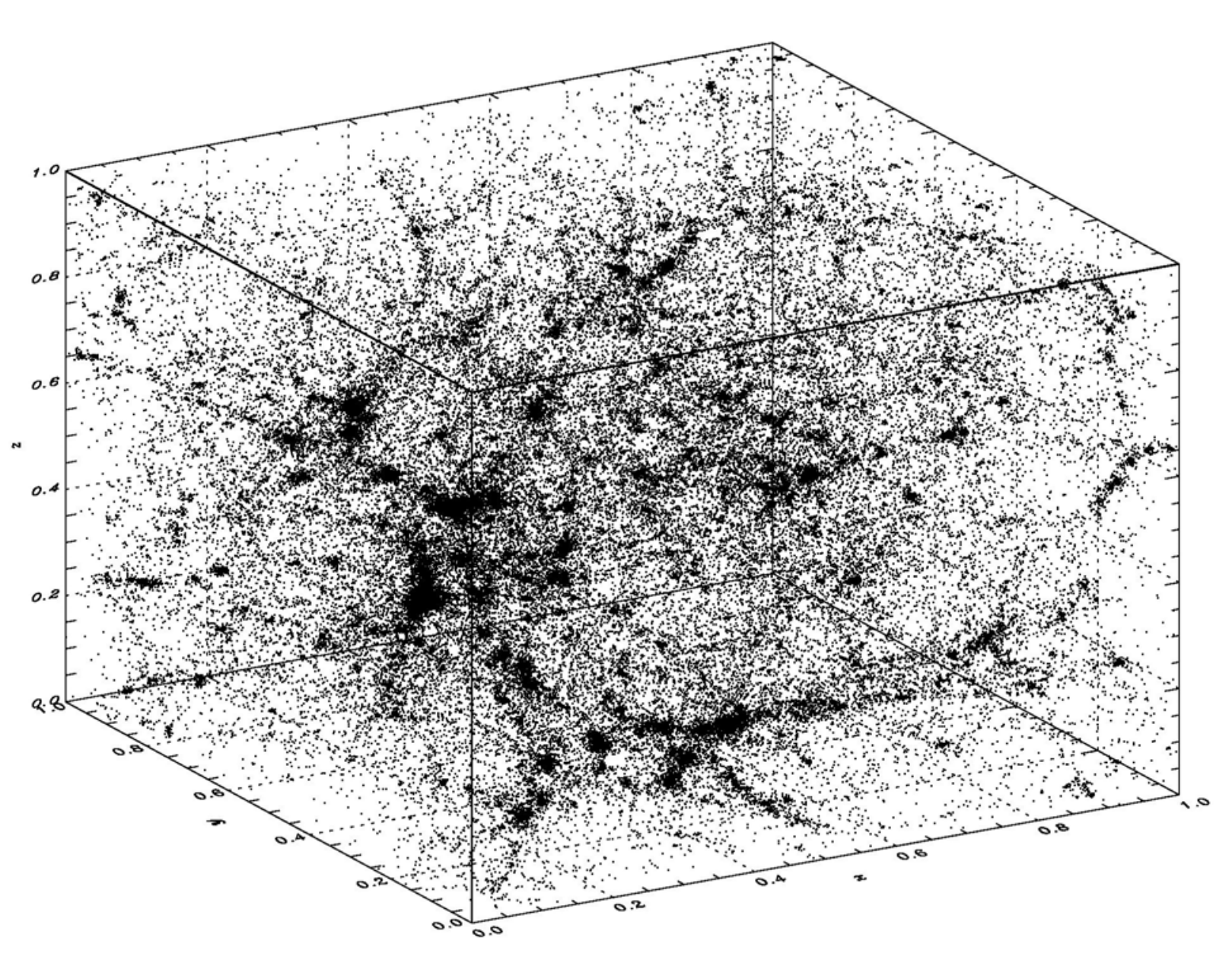}
\includegraphics[width=0.32\textwidth]{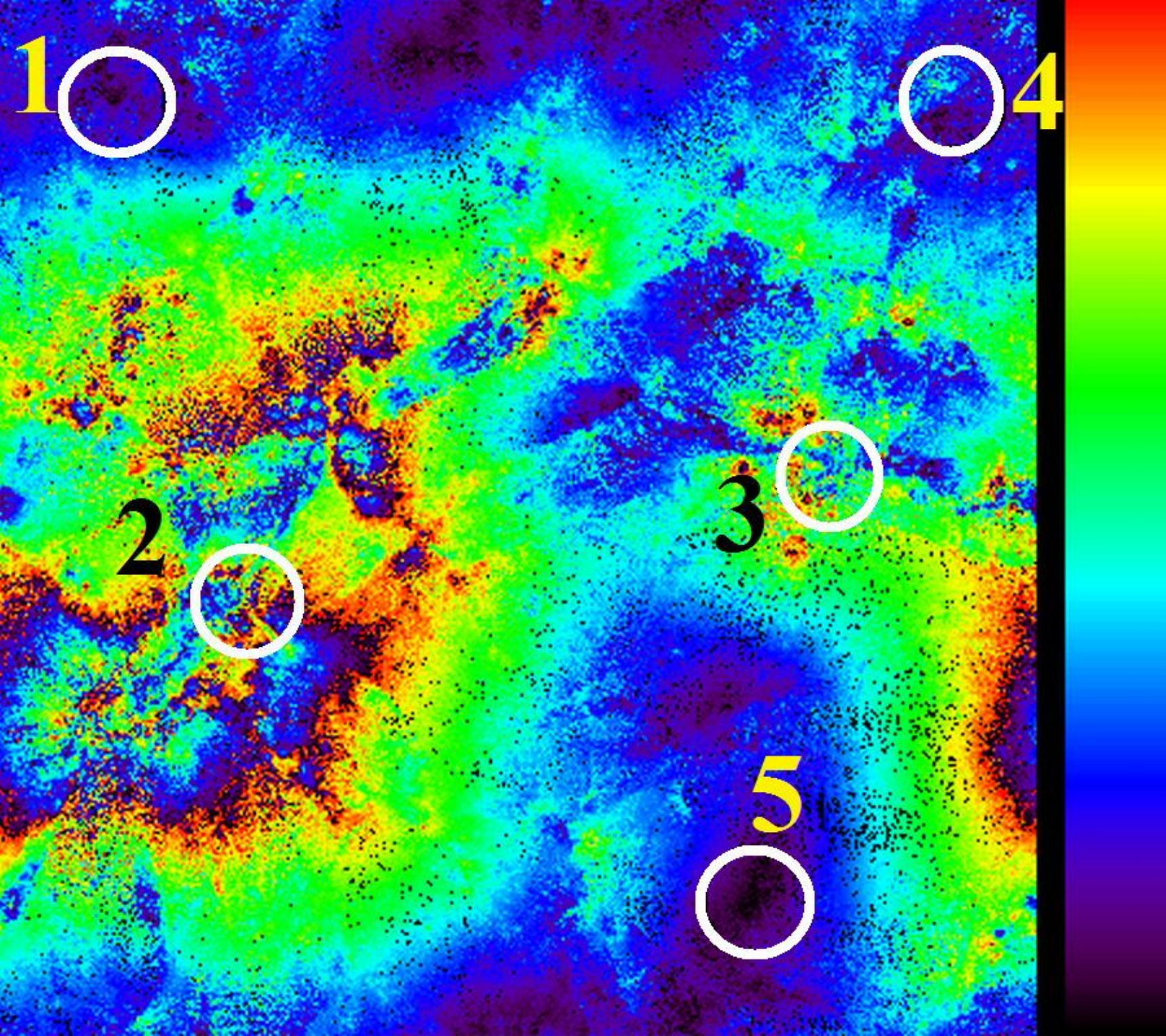}
\includegraphics[width=0.32\textwidth]{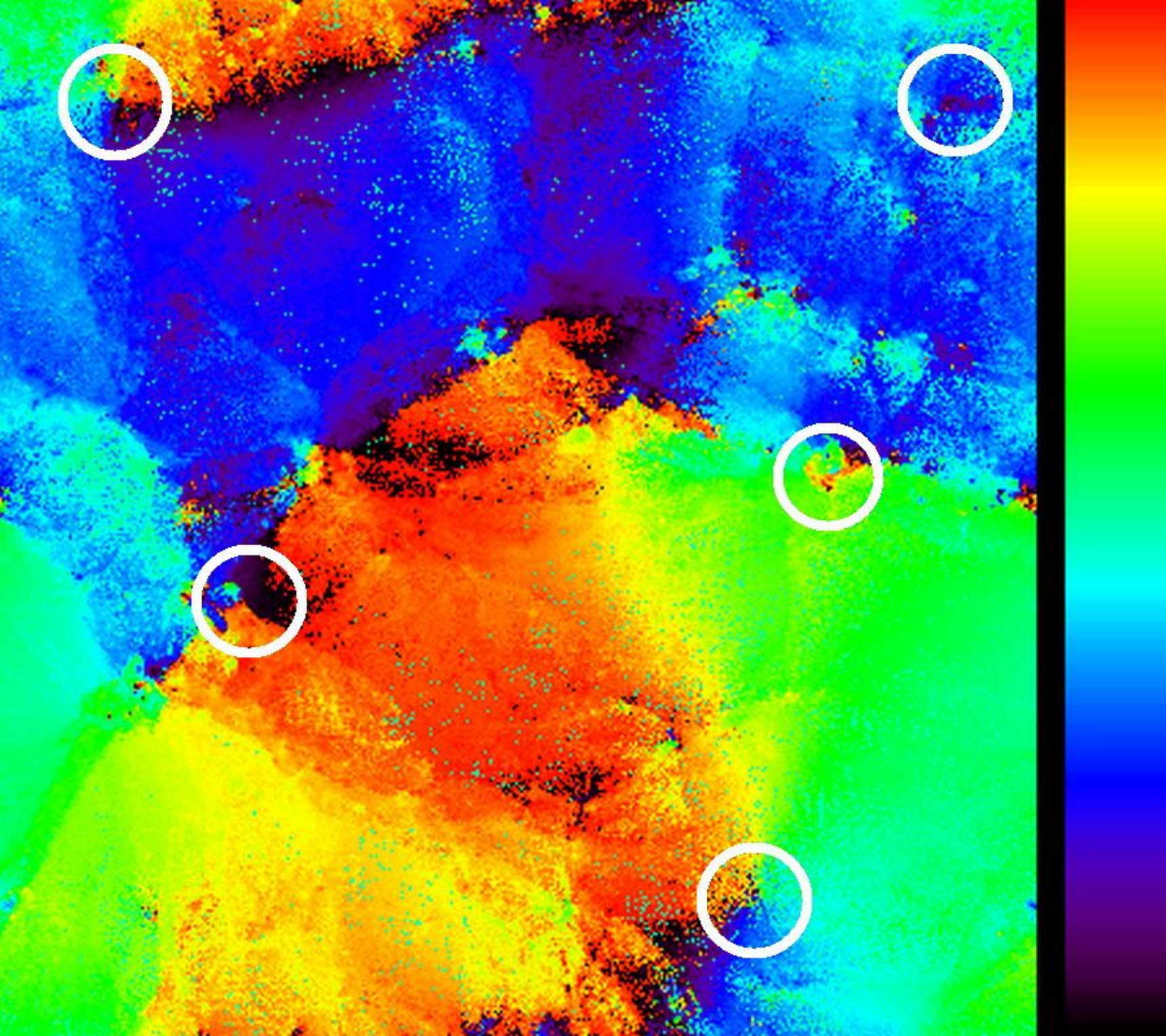}}
\caption{ Distribution of the density points in comoving coordinates (the left column) versus redshift from $z+1=50$ (top) down to $z=0$ (left bottom panel).   The middle and the right columns are for 2D cross-section
of the 3D map for the amplitude $|V|$ and the velocity angle $\Theta_v$ at the redshift $z=10,1.5,0.5,0$. All the color figures are taking
from the plane at the vertical axes $Z_c=0.5\pm |\Delta Z|$ with $|\Delta Z=0.1$ (see the next Section for details)  .
Five circles marks the position of the voids on the maps of the amplitudes and the velocity angle.
The color scale for $|v|$ (middle panels, dark purple to red) corresponds to the range of variability $[0,|v_{max}(z)|]$, and the color scale for the velocity angle (right panels, dark purple to red) corresponds to $[0,2\pi]$.
}
%\end{center}
\label{fig14}
\end{figure*}

In this section we discuss the evolution of the velocities around  the critical points,
using cosmological N-body simulations. We have used $256\times256\times256$ particles for a box of the linear size $85$ Mpcs with cosmological parameters given by Planck 2015 \cite{planck2015}.
Each simulation particle roughly represents a dwarf galaxy of mass $10^9$ $m_\odot$.

For the further analysis, we will use two dimensional cross-sections of the 3D velocity field. In the case when this cross-section is taking through the critical point at
 $Z_c=const$ -coordinate, the 3D velocity field in the vicinity of that  point would be represented by two dimensional vector ${\textbf v} = (v_x,v_y)$ (see Appendix for representation of the critical points in 2D.)
We would  like to introduce the following mapping technique for representation  of the velocity flow around the critical points, based on the absolute value of the vector ${\textbf v} = (v_x,v_y)$
and the vector angle:
\begin{eqnarray}
|\textbf {v}|^2=v^2_x+v^2_y, \hspace{0.2cm} \Theta=Atan(v_y,v_x)
\label{mod}
\end{eqnarray}
We have used already this kind of representation in Fig.~\ref{fig0}. In Fig.~\ref{histogram-velocity} we combine both vector and amplitude
representation for illustration of the properties of the velocity flow, given by
\begin{eqnarray}
 v(X,Y)=\frac{1}{g}\int_{Z_c-\Delta Z}^{Z_c+\Delta Z}v(X,Y,Z)dZ, \hspace{0.5cm} g=2\Delta Z
\label{cross}
\end{eqnarray}
 taking from the volume
$Z_c\pm \Delta Z, X,Y$ with the corresponding $Z_c=0.4125$ and the  width of the slice  $\Delta Z=0.0025$ . Here the length of $Z$ axis is normalized  to  85 Mpc , so $Z=[0,1]$.
We will call this representation as a
2D slice in Z-direction of the 3D distribution.  Firstly, in Fig.~\ref{histogram-velocity} we are focusing on the histogram
(the left panel) , which illustrate the relative concentration of the low velocity domains in respect to the high velocity patches.
This histogram can be nicely fitted by the log-normal  probability density function for $|v|$ with the following parameters:
\begin{eqnarray}
f(|v|)=\frac{1}{\sqrt{2\pi |v|\sigma}}\exp\left[-\frac{(\ln |v|-\mu)^2}{2\sigma^2}\right]
\label{log}
\end{eqnarray}
where:  $\mu=5.745$ and $\sigma=0.674$, and all the parameters are dimensionless.

The first moment of  the velocity distribution is  $<|v|>\sim 392$km/s and  the  second one is: $[<v^2>-<|v|>^2]^{\frac{1}{2}}\simeq 493$km/s.
At the same time,
the $|v|\le 50$ km/s domain corresponds to the 4$\%$ of the volume, while for $|v|\le 150$ km/s the corresponding fraction is 24\%.

 The  middle panel of
Fig.~\ref{histogram-velocity} show the vector representation of the velocity flow  with the length of the vector $|v|$ and direction, given by
the velocity angle from eqn.~\ref{mod}. The colored magenta vectors are pointed to the zones, where $|v|<50$ km/s. The right panel of this figure illustrate  a colored contour line representation of
the spatial distribution of the amplitude $|v|$. One can clearly see the correlations between the middle and the right plots, where the dark blue zones indicate the position of the voids. %In Fig.~\ref{histogram-velocity} we show the distribution of the velocity flow

As it was pointed out in the previous Section, for the zones around the critical points with $|v|=0$, the comoving coordinates $\chi_c$ should be close to
the coordinates $q_c$. Based on the N-body simulations, we would like to illustrate this peculiarity, presenting 3D evolution of the density field and the two dimensional amplitude- velocity angle representation from eqn.~\ref{mod} with redshifts $1+z=[1, 50]$. We fix the threshold $Z=0.5$  with the width $\Delta Z=\pm0.1$.
In this 2D cross-section we mark  the critical zones, mentioned in the previous section
and got the amplitude-velocity angle representation, shown in Fig.~\ref{fig14}. From  Fig.~\ref{fig0} one can find many points with
$|\vec{v}|\rightarrow 0$, which can be divided into three major groups. The first one corresponds to the
well established over-dense regions, associated with ``pancakes`` (see the black zones in Fig.~\ref{fig14}). The  others correspond to the voids and isolated points. As one can see from this figure, the morphology of the amplitude-velocity angle map is extremely stable and evolves little with time. This confirms that, due mainly to low amplitude of peculiar velocities at the critical zones, our classification
of the properties of the velocity potential at the linear stage of gravitational instability holds
well into the non-linear regime. \\

Next, we shall study the subsequent of our subsampling around the critical points for the variance of the Hubble flow $\sigma^2_H$.
From the analytical description we may expect that focusing on  those zones we can achieve the point of minima for $\sigma^2_H$. For illustration of this statement, we have taking under consideration five ``critical zones'', marked
by the white circles in Fig.~\ref{fig14}. Note that some of these zones correspond to well-developed structures, like a planes and
filaments with a void in between (zone 2). The others (zones 1 and 5) represent a critical stationary point like a void, zone 3 corresponds to a node or a cluster, and zone 4 is a saddle point of the velocity field, where the influence of
the large-scale structures is less pronounced (see Appendix for details.) We have identified 3D velocity components for these zones and
get the mean value of $\textbf{v}$, the standard deviation (r.m.s.) and $\sigma_H$ in the volume with radius $\varrho=4.25$ Mpc, presented
in Table~\ref{table1}:
\begin{table}[!htb]
\centering
\caption{The mean value and standard deviation of the velocity field for the whole volume and for the
zones 1-5 in Fig.~\ref{fig14} (the first and the second rows from the top).
All the amplitudes are normalized to 6223 km/s. The radius of the zones 1-5 is $\varrho=4.25$ Mpc.
The bottom row gives $\sigma_H$ for  the corresponding zones. }
\begin{tabular}{|c|c|c|c|c|c|c|}
\hline
     & Whole$\quad$ & zone 1$\quad$ & zone 2 $\quad$ & zone 3 $\quad$ & zone 4 $\quad$ & zone5 \\
\hline
   Mean &0.063 & 0.007&  0.047 & 0.063& 0.018& 0.011\\
   rms  &0.044 & 0.015& 0.061 & 0.019&0.034& 0.006\\
$\sigma_H$&0.093&0.014&0.057&0.017&0.031&0.0056\\
\hline
\end{tabular}
\label{table1}
\end{table}

As it is follows from Table~\ref{table1}, for the cluster (zone 3) the mean velocity and the velocity within the radius
$\varrho\simeq 5$ Mpc are almost the same, while rms are smaller by a factor of about 2.3. For the two voids (zone 1 and 5) the mean and the rms values are smaller than for the whole volume by a factor of $\sim(5.7-9)$ (mean), and
$(2.9- 7.3)$(for rms). For the saddle point the corresponding ratios are $3.5$ (mean) and $1.3$(rms). As it follows from eqn.~\ref{vec3} the corresponding $\sigma_H$ for the $R=L=42.5$ Mpc and $\varrho\simeq 4.25 $Mpc is presented in Table 1 on the third row. Note that for the whole volume (on the left of the third row), the
$\sigma_H$ value is in full agreement with the result given in \cite{wang98}. Moreover, in the vicinity of the stationary points (zones 1 and 5), the corresponding $\sigma_H$ is about 6-16 times smaller than for the whole volume , as it was predicted in Section 1. Taking into account, that $\sigma_H$ in zones 3 and 5 is proportional to
the size $\varrho$, even extension of the volume around these stationary points up to 18 Mpc will make
$\sigma_H$ still four time smaller than for the whole volume (with $R=L$).

\section{Conclusion}
%-----------------------------

The formation of structures on large scales has been well-described by the Zel'dovich approximation which
predicts the formation of the observed walls, filaments and nodes in the cosmic density field. The vanishing of the
Jacobian of the transformation between Lagrangian and Eulerian coordinates formulates the formation of such formally singular structures. The Jacobian is basically the Hessian of the gravitational potential.
In this work, we aim at extending the previous works on the density field to the cosmic velocity field and show how one can search systematically for the vanishing zones in the cosmic field of peculiar velocities. These zones are interesting for various reasons. Peculiar velocities, although extremely informative on various problems such as the quantity and distribution of mass, the value of the bias parameter etc, are somehow nuisances for the determination of certain cosmological parameters, such as the Hubble constant or dark energy from various observations and in particular SNe Ia and  baryonic acoustic oscillation (BAO) \cite{seo2010}. Typically, peculiar velocities increase the scatter in the dark-energy parameter and displace and or wash-out the peaks of BAO. Since distances and peculiar velocities cannot be measured independently,
It is extremely hard to directly and precisely estimate the effect of peculiar velocities and one often uses simulations to infer knowledge of the amount of the resulting scatter.

In this work, we have shown how to reduce systematically the errors in such parameter-estimations by restricting one's analysis within zones of vanishing peculiar velocities instead of averaging and using the entire field  which is the usual procedure.  We have shown how these zones can be identified on large scales.
The velocity field remains a potential flow well into the non-linear regime and up to third order in the Lagrangian perturbation theory, and follows the gravitational potential even in the mildly nonlinear regime. The vanishing zones of the velocity field correspond to the vanishing of the first derivative of the gravitational potential. The nature of these zones can be classified by analyzing the Hessian of the gravitational potential. The maxima, minima correspond to the voids, the center of dense structures such as clusters and the saddle points correspond to walls and filaments. Our studies of N-body simulations show that the zero-velocity zones are mainly in the void regions. As these regions are static, they can retain the memory of the initial conditions and hence can be used to trace back and recover the primordial distribution of matter with little interference from the peculiar velocities.
Using specific toy models of potential, we have tried to show how to identify and classify these zones. we have then tested our method against $\Lambda$CDM N-body simulations and have shown that when searching for the value of the Hubble parameter, one can reduce the variance significantly by taking a subsample in and around critical points where the first derivative of the gravitational potential is zero rather than using the whole simulation box. The application of our method to observational data shall follow in subsequent works.

\section{Acknowledgments}
RM acknowledges support from Nordea foundation and grant from DNRF Niels Bohr Professorship awarded to Subir Sarkar. We thank Subir Sarkar and Sergei Shandarin for discussions, Adam Riess, Giovanni Marozzi and Per Rex Christensen for useful comments, and Stephane Rouberol for maintaining
the Horizon cluster at IAP where a few of our simulations were run.
This work is supported in part by Danmarks Grundforskningsfond, which allowed the establishment of the Danish Discovery Center.
This work is supported by FNU grant 272-06-0417, 272-07-0528 and 21-04-0355 and Villum fond through
Deep Space project.
Hao Liu is supported by the National Natural Science Foundation of China (Grant No. 11033003), the National Natural Science Foundation for Young Scientists of China (Grant No. 11203024) and the Youth Innovation Promotion Association, CAS.

\section{References}
%\begin{references}

\section{Appendix}

 \subsection{Critical points of the cosmic velocity field in 2D.}

Projection of the 3D velocity field on the plane, given by eqn.~\ref{cross}, allow us to classify the properties of the critical points in the same way, as for the system of linear differential equations\cite{bronshtein} .Since we are interested in the behavior
 of the cosmic velocity field in the vicinity of the critical points, {\it i.e.} where $|v_i|\rightarrow 0$, the linear approximation (see  eqn.~\ref{eqV}) still holds.
 The velocity potential  in the linear regime,  $\Phi(\textbf{q})$  around the points of minima or  maxima is given by
%   At the vicinity of the extreme and the saddle points,  tFor the points of extreme  substitution of
    eqn.~\ref{eqEX} and  eqn.~\ref{eqV}, which leads to the system of linear equations,
\begin{eqnarray}
 \frac{d\Delta \chi_i}{d\eta}=\sum_k\alpha_{ik}\Delta\chi_k\,.
\label{sys}
\end{eqnarray}
We shall search for the solution of  eqn.~\ref{sys} in the form of $\Delta\chi_k=A_ke^{\lambda\eta}$, where $A_k$ is the amplitude and the parameter $\lambda$ satisfies
%\textcolor{red}{This is not very clear ? why ? can we show this more pedagogically please ?}
the following  equation:
\begin{eqnarray}
 {\rm det}(\alpha_{ij}-\lambda\delta_{ij})=0\,,
\label{eq11}
\end{eqnarray}
where $\delta_{ij}$ is the Kronecker symbol.
Thus, the $\lambda$-parameter and the vector $A_k$ are the eigenvalues and eigenvector of the deformation tensor at the points of extreme. It is well known that by linear transform of the coordinates
$\Delta\chi_i=\sum_jC_{ij}\varsigma_j$, where $C_{ij}$ is a matrix and $\varsigma_i$ the new coordinates, the potential $\Phi(\varsigma_i)$ of  eqn.~\ref{eqEX} can be expressed as:
\begin{eqnarray}
 \Phi(\varsigma_j) =\Phi_0(0)+\frac{1}{2}\sum_i\lambda_i\varsigma^2_i\,.
\label{norm}
\end{eqnarray}
If all $\lambda_i$ are positive, we have a point of minima of the velocity potential at $\varsigma_j=0$, and if all  $\lambda_i<0$,
the point $\varsigma_j=0$ is a point of maximum.
\begin{figure*}[!htb]
 % \begin{center}
\hbox{
\centerline{\includegraphics[scale=2.6]{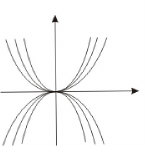}
\includegraphics[scale=2.6]{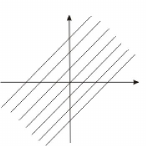}%}%}
%    \centerline{
\includegraphics[scale=2.6]{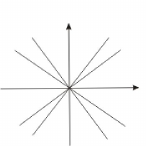}
\includegraphics[scale=2.6]{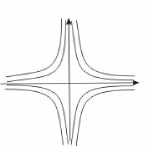}}}
   \caption{ The structure of the velocity field in the vicinity of the critical points of the gravitational potential, called \textit{nodes}, with $\lambda_1\neq \lambda_2$, $\lambda_1\lambda_2>0$ , both $\lambda_{1,2}$ are real (the left panel).
The second from the left panel corresponds to  $\lambda_1=0, \lambda_2\neq 0$,  both $\lambda_{1,2}$ are real .
 The third from the left is for $\lambda_1= \lambda_2\neq 0$,  both $\lambda_{1,2}$ are real . The right panel show the \textit{saddle point} of the velocity field (but not the gravitational potential !)
with  $\lambda_1\neq \lambda_2$, $\lambda_1\lambda_2<0$ , both $\lambda_{1,2}$ are real.
}
    \label{figA}
%  \end{center}
\end{figure*}

In Fig.~\ref{figA} we plot the typical structure of the velocity field for real $\lambda_{1,2}$ . One can see that all the features, presented in
Fig.~\ref{histogram-velocity} and Fig.~\ref{fig14} can be easily understand   in terms of the critical points of the linear differential equations\cite{bronshtein}.
\end{document}